\documentclass[10pt,twocolumn]{article}
\usepackage{times}
\usepackage{cite}
\usepackage[pdftex]{graphicx}
\usepackage{epsfig}
\usepackage{url}
\usepackage{multirow}
\usepackage{xspace}
\usepackage{color}
\usepackage{breakurl}
\usepackage[breaklinks]{hyperref}

\begin{document}
\newcommand{\itempar}[1]{ \noindent \textbf{#1}}
\newcommand{\mysec}[1]{\section{#1}\vspace{-0.1in}}
\newcommand{\mypar}[1]{\vspace{-0.02cm} \noindent \textbf{#1:}}
\newcommand{\myitem}{\vspace{-0.1cm} \item}
\newcommand{\todo}[1]{\textbf{TODO: [#1]}}
\newcommand{\TODO}[1]{\textbf{TODO: [#1]}}
\newcommand{\fixme}[1]{\textbf{FIXME: [#1]}}
\newcommand{\Fixme}{{\bf FIXME:}}
\newcommand{\remark}[1]{\textbf{REMARK [#1]???}} 
\newcommand{\response}[1]{\textbf{RESPONSE [#1]???}} 
\newcommand{\dropme}[1]{\textbf{DROPME [#1]???}} 
\newcommand{\reviewcomment}[2]{\textbf{REVIEW \##1}{\xspace\emph{#2}}} 
\newcommand{\eg}{\textit{e.g.}}
\newcommand{\ie}{\textit{i.e.}}
\newcommand{\etal}{\textit{et al.}\xspace}
\newcommand{\codesm}[1]{\texttt{\small #1}}
\newcommand{\ignore}[1]{}
\newcommand{\budget}[1]{\textbf{SPACE BUDGET: #1}}
\newcommand{\myincludegraphics}[2]{\resizebox{#1}{!}{\includegraphics{#2}}}
\newcommand{\red}[1]{{\color{red} #1}}

\long\def\manishc#1{[{\color{blue}Manish writes: \it #1}]}
\long\def\mijungc#1{[{\color{red}Mijung writes: \it #1}]}

\newcommand*\BitAnd{\mathrel{\&}}
\newcommand*\BitOr{\mathrel{|}}
\newcommand*\ShiftLeft{\ll}
\newcommand*\ShiftRight{\gg}
\newcommand*\BitNeg{\ensuremath{\mathord{\sim}}}


\title{Sparkle: Optimizing Spark \\ for Large Memory Machines and Analytics}
\author{Mijung Kim$^1$, Jun Li$^1$\thanks{Currently at Ebay},  Haris Volos$^1$,  Manish Marwah$^1$, Alexander Ulanov$^1$\thanks{Currently at Facebook}, \\ 
Kimberly Keeton$^1$, Joseph Tucek$^1$\thanks{Currently at Amazon Web Services}, Lucy Cherkasova$^1$\thanks{Currently at HyTrust}, Le Xu$^2$, Pradeep Fernando$^3$ \\
\small {\em  $^1$ Hewlett Packard Labs } \\
 \small  {\em $^2$University of Illinois \quad 
           $^3$Georgia Institute of Technology} \\ [2mm]
\small Submission Type: Research
}
\date{}
\maketitle

\begin{abstract}
Spark is an in-memory analytics platform that targets commodity server environments today.
It relies on the Hadoop Distributed File System (HDFS) to persist intermediate checkpoint states and final processing results.
In Spark, immutable data are used for storing data updates in each iteration, making it inefficient for long running, iterative workloads. 
A non-deterministic garbage collector further worsens this problem.
Sparkle is a library that optimizes memory usage in Spark.
It exploits large shared memory to achieve better data shuffling and intermediate storage.
Sparkle replaces the current TCP/IP-based shuffle with a shared memory approach and proposes an off-heap memory store for efficient updates.
We performed a series of experiments on scale-out clusters and scale-up machines.
The optimized shuffle engine leveraging shared memory provides 1.3x to 6x faster performance relative to Vanilla Spark. 
The off-heap memory store along with the shared-memory shuffle engine provides more than 20x performance increase on a probabilistic graph processing workload that uses a large-scale real-world hyperlink graph.
While Sparkle benefits at most from running on large memory machines, it also achieves 1.6x to 5x performance improvements over scale out cluster with equivalent hardware setting.
\end{abstract}

\section{Introduction}

Apache Spark \cite{zaharia2016cacm} is perhaps the most popular large scale 
data processing framework available today.
Its popularity stems primarily from its capability for in-memory 
fault-tolerant computation on large-scale commodity clusters and support
for a broad range of data processing and analytics applications, including 
SQL processing, machine learning, stream processing, and graph analytics.
Spark, similar to other large-scale analytics frameworks~\cite{dean:mapreduce:osdi:2004, hadoop},
is designed for a cluster of commodity low-end server machines following 
a scale-out architecture. 
Scale-out clusters are attractive as they enable such frameworks to crunch 
huge datasets in a cost-effective manner.

However, recently large memory servers equipped with 100s of GBs of DRAM 
and tens of cores, also known as scale-up systems, have become more available 
and affordable.
For example, a 96 core, 1 TB machine is available for under \$38K~\cite{compsource}.
In fact, cloud service providers such as Amazon Web Services (AWS) now offer 
large memory instances with up to 1,952 GB of DRAM and 128 vCPUs\cite{amazon-x1}.
Such scale-up systems are attractive since they can be provisioned and 
optimized for better performance compared to a cluster, especially when 
the workload is not embarrassingly parallel. 
As we show later in the evaluation (Section \ref{sec:experiments}), for 
equivalent hardware configuration, a scale-up implementation provides
at least 1.6x up to more than 5x speed-up over a scale-out system, 
with scale-up settings being especially beneficial to communication 
intensive workloads and large-scale iterative workloads in our 
experiments.

Given the growing availability of affordable scale-up servers,
our goal is to bring the performance benefits of in-memory processing 
on scale-up servers to an increasingly common class of data analytics 
applications that process small to medium size datasets (up to a few 100GBs)
that can easily fit in the memory of a typical scale-up server~\cite{appuswamy2013scale}.
To achieve this, we choose to leverage Spark, an existing memory-centric 
data analytics framework with wide-spread adoption among data scientists.
Bringing Spark's data analytic capabilities to a scale-up system
requires rethinking the original design assumptions,
which although effective for a scale-out system, are a poor match 
to a scale-up system resulting in unnecessary communication and memory 
inefficiencies.

\ignore{
 Spark is an existing dataflow engine for large-scale in-memory data processing that targets commodity server cluster environments.
 One critical performance path in Spark (and other dataflow processing engines such as Hadoop) is data movement, both in initial load and in shuffling.
 To load data, Spark relies on HDFS, which is a disk-centric shared-nothing distributed file system implemented by tying individual nodes’ local file systems together with a remote-access layer.
 This approach introduces unnecessary layering overheads that hide the shared memory performance.
 To shuffle data between the nodes of a cluster, Spark today relies on TCP/IP, which introduces unnecessary (and non-trivial) overheads in a shared memory environment.
 Furthermore, Spark’s shuffle engine is implemented in Scala running in a Java Virtual Machine (JVM) environment; as a result, shuffling a large amount of data requires creating a large number of Java heap objects, putting considerable pressure on the JVM’s garbage collector.
}

Since Spark targets commodity scale-out clusters, it follows a shared-nothing 
architecture where data is partitioned across the nodes of a cluster. 
Spark hides partitioning and distribution through the concept of 
\emph{resilient distributed datasets} (RDDs), an immutable collection 
of partitioned data created through deterministic transformations on data.
Although most transformations on partitions can be performed independently by 
each node, several transformations such as joins and group-by require global 
communication to shuffle data between worker nodes, which Spark performs through 
the TCP/IP networking stack. 
While effective on scale-out systems interconnected with Ethernet, the networking 
stack introduces unnecessary copying and serialization/deserialization overheads 
in a scale-up system where communication can be performed through shared memory instead.

As Spark is implemented in Scala, a memory managed language based on the Java 
Virtual Machine (JVM), this leads to memory inefficiencies due 
to increased memory pressure put on the garbage collector (GC) by a large 
number of heap objects created when shuffling data, and immutable RDDs.
Spark leverages immutability of RDDs to support fault tolerant 
computation by tracking how RDDs are constructed from other RDDs (lineage) 
and recomputing RDDs upon failure.
This strategy, while generally effective, can hinder the scalability of 
Spark with memory intensive, iterative workloads, such as those common 
in large-scale machine learning applications (\eg, gradient descent on a 
large data set) and large graphs analytics, which create and cache many 
RDD instances in managed memory, thus increasing memory consumption 
and putting pressure on the GC.
\ignore{
 As garbage collection timing is non-deterministic, even unreferenced 
 RDDs continue to reside in managed memory until the next GC invocation.
}
To reduce pressure on the GC, Spark may evict cached RDDs. However, 
evicted RDD partitions have to be re-computed and re-cached when needed 
for the next iteration, which further evicts currently cached partitions. 
Eventually the re-computation can snowball, impeding any further progress 
on the current iteration.

\ignore{
 \item Immutable RDD: The RDDs are immutable and are used to store data
  updated in each iteration, thus requiring multiple instances of
  cached RDDs and corresponding increase in memory consumption.
 \item Non-deterministic garbage collection: The Garbage Collection
  (GC) timing is non-deterministic. Unreferenced RDDs continue to
  reside in heap memory until next GC invocation, further increasing
  memory consumption.
 \item RDD cache eviction policy: Heap memory is often not enough to
  accommodate these unneeded RDDs in cache, resulting in eviction of
  some cached RDD partitions. The evicted RDD partitions have to be
  re-computed and re-cached when needed for the next iteration, which
  further evicts currently cached partitions. Eventually the
  re-computation snowballs, impeding any further progress on the
  current iteration.
}

To address the inefficiencies and scalability issues described above, 
we have designed and implemented {\em Sparkle}, an enhancement of Spark 
that leverages the large shared memory available in scale-up systems to 
optimize Spark's performance for communication and memory intensive 
iterative workloads. Further, we have open-sourced Sparkle~\cite{sparkle}.

Specifically, our work makes the following contributions in addressing 
the above challenges:

\begin{itemize}
  \item A scalable shared-memory management scheme and allocator that 
allows worker processes to efficiently store data in native global shared memory
without incurring the overhead of Java managed object representation. 
  \item A shared-memory shuffle engine where shuffle map and reduce tasks 
communicate through references to global shared memory without incurring 
serialization/deserialization and copying overheads associated with communication 
through the networking stack. 
  \item An off-heap memory store that allows data caching in global shared 
memory to reduce the pressure put on managed memory by the large number of RDDs 
typical in large-scale iterative workloads.
  \item A thorough evaluation that shows Sparkle's advantages compared 
against original Vanilla Spark on both scale-up and scale-out settings on a 
variety of workloads. Notably, Sparkle achieves 1.6x to 5x performance 
improvements over Spark running on scale out cluster with equivalent hardware 
setting. 

\end{itemize}

\ignore{
The rest of the paper is organized as follows.
Section \ref{sec:relatedwork} describes related work.
Detailed description Sparkle and its main components are presented in Section \ref{sec:methodology},
followed by a section on experiments and performance evaluation.
Finally, the conclusion are presented in Section \ref{sec:conclusions}.
}

\section{Related Work}
\label{sec:relatedwork}
There has been a plethora of previous work that looked at optimizing the 
performance of Spark through various techniques. 
Below we review efforts that we find are most closely related.

\paragraph{Optimizing for scale-up machines.}
Appuswamy \etal explored the question of scale-up vs scale-out in the 
context of Hadoop~\cite{appuswamy2013scale}.
They find that the scale-up approach can offer competitive and often 
better performance than scale-out for several workloads that can fit 
in the memory of a single server machine. 
To achieve this they propose several modifications to the Hadoop 
runtime that require no application changes. This includes 
optimizing the shuffle phase to transfer data by writing 
and reading to a local file system rather than copying data 
over the network. 
We initially considered this approach but quickly abandoned it in 
favor of our in-memory shuffle engine.
The file system approach offered limited performance improvements 
due to contention inside the operating system kernel especially on 
large-scale machines such as Superdome X and also suffered from 
copying and serialization/deserialization overheads similar to 
the default networking approach.

\paragraph{Optimizing for HPC systems.}
Previous work has explored optimizing and tuning Spark for high-performance 
computing (HPC) systems. HPC systems pose several differences from systems 
found in data center environments, such as communicating through high-speed 
RDMA-enabled interconnects and storing all data to a global parallel file 
system like Lustre~\cite{lustre} due to lack of locally attached storage.
Lu \etal accelerate the shuffle phase in Spark by leveraging RDMA to avoid 
the overhead of socket-based communication~\cite{rdmaspark}.
Chaimov \etal find that placing a large NVRAM buffer pool between compute 
nodes and the Lustre file system helps improve scalability of Spark running 
on a Cray XC machine~\cite{chaimov2016scaling}.
\ignore{find that running Spark efficiently on a Cray XC machine 
requires extending each compute node with a local file system to cache the 
intermediate results produced during the shuffle rather than storing them 
in Lustre \cite{chaimov2016scaling}.}
Although these optimizations are less applicable to the data center today, 
they may become handy as the data center begins to adopt technologies 
like RDMA over Converged Ethernet (RoCE)~\cite{roce}.

\paragraph{Exploiting native memory.}
Project Tungsten from Databricks~\cite{tungsten} aims to improve Spark's JVM memory 
management by pushing data structures into off-heap native memory via Java Unsafe APIs, 
thus reducing garbage collection overhead, as well as exploring code generation based 
on schema definitions.
Most efforts to date have been devoted to improving Spark SQL’s DataFrame 
based processing.
FACADE provides a compiler pass that can transform existing data-centric 
applications, including Spark, to store objects in native off-heap memory 
with minimal application modifications required by the programmer~\cite{nguyen:facade:asplos:2015}.
Overall, our approach of completely re-writing the entire shuffle engine 
and introducing an off-heap store based on shared memory 
goes beyond laying out native in-memory data structures in individual JVM 
processes to more fully exploit the shared-memory pool.

\paragraph{Leveraging specialized data stores and formats.}
Key-value stores have been explored to scale machine learning algorithms such as gradient descent~\cite{li2014scaling}. 
Most existing approaches rely on TCP/IP based methods for remote fetching of attributes.
In our off-heap memory store, attributes (local or remote) are stored in native data structure layout and accessed directly via shared-memory, which shortens access latency and improves bandwidth utilization for bulk attribute access.

Spark DataFrame based processing has been extended to graph processing, called GraphFrames~\cite{dave2016graphframes}, with focus to support graph query processing.
In DataFrames, the off-heap data structures are designed to represent to-be-queried data in a columnar format and to speed up local query processing.
These data structures are local and private to individual Spark executors.
In contrast, our off-heap memory store supports global direct memory access on the constructed data structures, which allows us to reduce shuffle stages for certain algorithm implementations.
DataFrames inherits RDD’s immutability and thus processing models represented in DataFrames is not updatable, whereas models stored in our attribute tables are updatable.
It is also worth mentioning that since our globally visible data structures are constructed from shared-memory, these data structures survive even after all Spark executors are terminated when a Spark job is finished, and they can be accessed across different Spark jobs, or even for non-Spark applications when necessary. 

In-memory storage systems support computing and transacting on large-scale data sets on in-memory platforms. Tachyon~\cite{li2014tachyon}  provides a persistent store on off-heap memory. Apache Ignite~\cite{ignite}  provides in-memory Shared RDDs for Spark jobs. Apache Arrow~\cite{arrow}  is specialized in a columnar memory-layout for fast random access. Most systems still suffer from data movement issues such as serialization/deserialization while our off-heap memory store provides in-situ data structures that are globally sharable.


\section{Sparkle Architecture}
\label{sec:methodology}

\begin{figure*} 
\psfig{figure=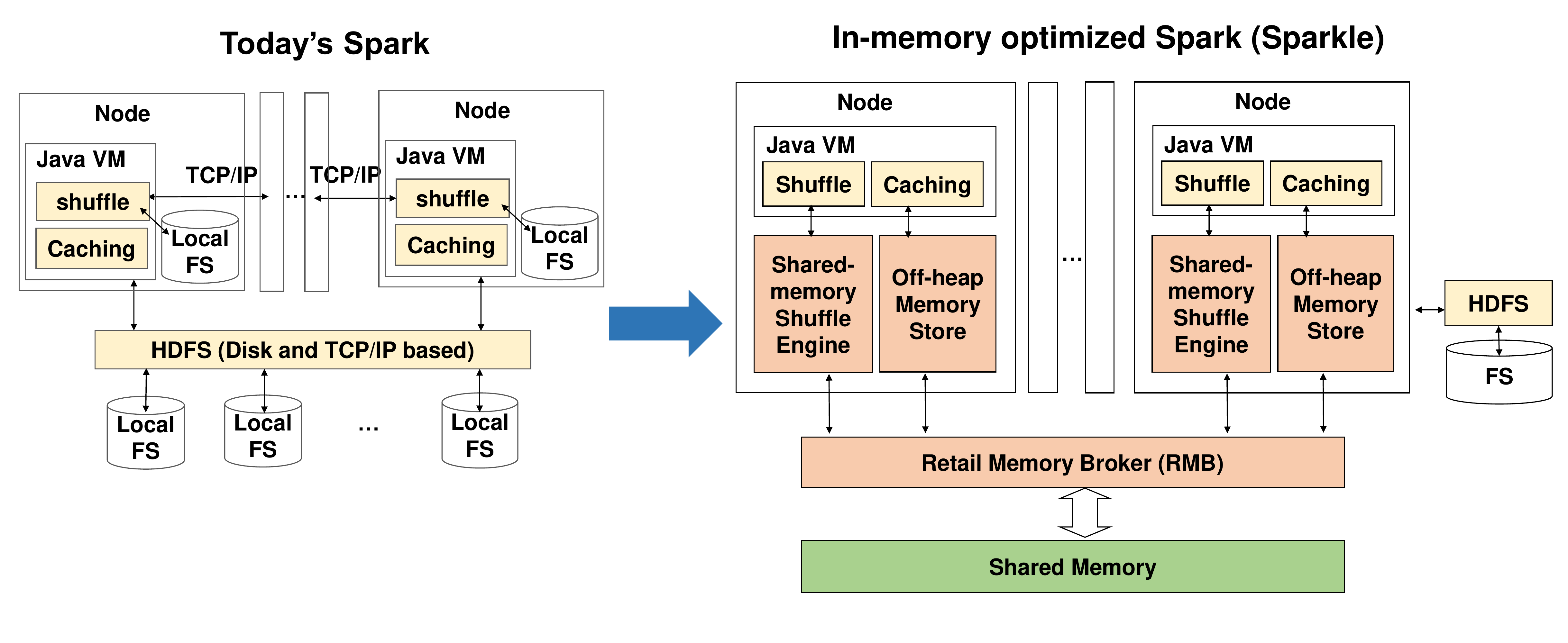, width=6.5in}
  \caption{Sparkle using the global shared memory-based architecture to transform Spark from a cluster-based scale-out architecture to a scale-up architecture}
  \label{fig:sparkle}
\end{figure*}

A key goal of Sparkle is to leverage the global shared memory architecture of
scale-up machines so that worker processes can share data in a memory 
and communication efficient manner. 
We start by briefly reviewing how Sparkle's system architecture achieves 
this goal, and then dive into each Sparkle component. 

\subsection{System Overview}
Figure~\ref{fig:sparkle} shows how Sparkle exploits global shared memory 
to transform Spark from a cluster-based scale-out architecture to a scale-up 
architecture.
At the bottom, a retail memory broker (RMB) layer provides a native memory management 
scheme that allows higher layers allocate and free blocks of global shared 
memory in a scalable manner. 
Worker processes can use native shared memory to store data without incurring 
the memory overhead of Java managed object representation and communicate without 
suffering the overheads of copying and serializing/deserializing data through
the networking stack.

\ignore{
 With respect to data shuffling, Sparkle replaces the traditional TCP/IP-based shuffle with a shared memory approach to write to and retrieve shuffle data from the shared memory. We aim to accelerate today’s data analytics applications by leveraging existing in-memory data processing frameworks. Spark is an existing dataflow engine for large-scale in-memory data processing that targets commodity server cluster environments. Spark, which is the most active Apache project in the big data area, has been widely used for various analytics tasks that include machine learning, streaming, SQL processing and graph processing. Although Spark focuses on in-memory processing, it still has many inefficiencies in terms of data shuffle and data caching.
}

For data shuffle, we have developed a shared-memory shuffle engine and 
integrated it into Spark under its pluggable shuffle interface. 
The shared-memory shuffle engine replaces the traditional TCP/IP-based 
shuffle with a shared memory approach to write to and retrieve shuffle 
data from the shared memory. 
Shuffle data is exchanged through shared-memory blocks managed by the 
retail memory broker rather than TCP/IP. 
Our solution is implemented in C++ for high performance, and it is
integrated into Spark through JNI (Java Native Interface), which is a 
mechanism that enables native C/C++ code to be called from within JVM 
programs.

For data caching, we have developed an off-heap memory store that allows 
us to construct various large scale data structures in shared-memory regions 
managed by the RMB. The data structures developed include a sorted array 
and a hash table, to store intermediate data processing models, and to allow 
these models to be updated in place during iterations. 
Moreover, since these data structures are globally accessible by every Spark 
executor within the cluster, we can reduce the number of shuffle stages 
that are required to implement the iterative processing related algorithms. 
This is achieved by having the data written (or updated) in shared memory in 
the current processing stage by a single writer, to be read directly from 
shared memory by multiple readers in the next processing stage. 
As data stored in the off-heap memory store are mutable, RDD lineage cannot
be used to reconstruct such data after a failure. To address this issue, the 
off-heap memory store provides fault tolerance through lightweight checkpoints 
that are taken at consistency points identified by the application developer.

The proposed off-heap memory store provides the following advantages: 
(1) mutable storage that can be used for iterative processing, thus resolving the problem of memory pressure from multiple immutable RDDs; 
(2) significantly reduced GC overhead; 
(3) more compact layout of attributes in memory store compared to Java objects; 
(4) globally shared data structures in off-heap memory store. 
These advantages allow processing of larger graph workloads with the same 
compute/memory resources. 

\subsection{Shared Memory Management}

\ignore{
 A key goal of our work is to leverage shared memory available in scale-up 
 machines to optimize interprocess communication between multiple processes
 running on the same machine. 
 Although processes could continue relying on high-level communication primitives
 such as network sockets that are agnostic to the underlying interconnect,
 such transparency comes at the cost of network stack software overheads 
 including extra memory copies needed to transfer data between processes.
 To avoid such overheads, we instead choose to communicate via
 references to global memory that is shared between multiple processes. 
 Under this model, a process that wishes to share data allocates a chunk of 
 global memory, stores data in the memory chunk, and then passes a reference 
 to the chunk to another process. The other process can use the reference to 
 locate and access data directly from global memory, without having to first 
 copy data into a local memory buffer. 
}

We provide a native memory management layer called the Retail Memory 
Broker (RMB) for use by higher-level system components, including the 
shuffle engine and the off-heap memory store.
The broker exposes a shared heap abstraction to global memory, with 
users being able to dynamically create and destroy multiple shared heap 
instances over the global memory pool. Each heap instance is identified
by a unique generation number. 

A user can allocate and free variable-size chunks of memory from a heap 
instance through a malloc/free-like interface, and pass references to 
allocated memory to other processes for shared access.
Although only a single owner process can allocate and free memory from 
a heap instance at a time, multiple processes can concurrently share 
and access memory associated with a heap.
Under this model, a process that wishes to share data allocates a chunk of 
global memory, stores data in the memory chunk, and then passes a reference 
to the chunk to another process. The other process can use the reference to 
locate and access data directly from global memory, without having to first 
copy data into a local memory buffer. 
The malloc method also allows users to supply an optional location hint, 
which is the socket node to allocate memory from.
This can be useful for processes that exhibit locality of reference 
such as shuffle map tasks that may wish to produce and store their 
output into physically close memory.

Destroying a heap instance frees up all memory chunks associated with 
that heap, which can be especially useful for collectively freeing 
up multiple objects that have a common lifetime. 
For example, intermediate objects created during the course of a shuffle 
stage all become dead and can be freed together when the stage completes. 
Supporting multiple heap instances gives us the benefit of efficiently 
managing memory objects with common lifetime similarly to region-based 
management~\cite{tofte:region-mem:jic:1997,gog:broom:hotos:2015}, while 
being able to manage memory of individual objects for better memory 
efficiency~\cite{berger:custom-mem-alloc:oopsla:2002} as needed by the 
use case of the off-heap memory store.

\begin{figure} 
\psfig{figure=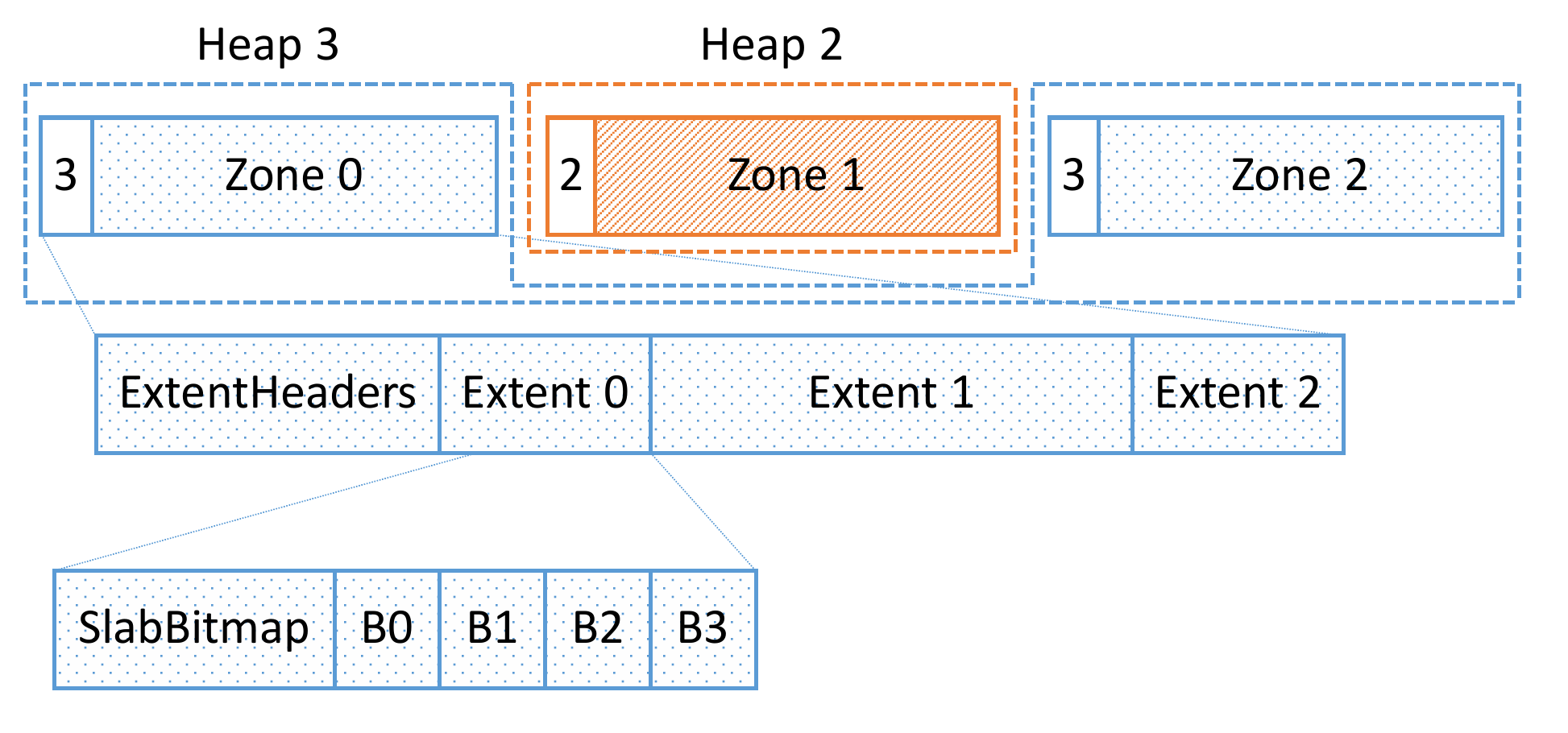, width=3.2in}
  \caption{Multiple shared heap instances overlayed over a global memory 
pool, which is partitioned into fixed-size zones. Heap 2 comprises zone 1, 
and Heap 3 comprises zones 0 and 2.}
  \label{fig:rmb}
\end{figure}

The broker internally organizes the global memory pool into \emph{zones}
as shown in Figure~\ref{fig:rmb}.
A zone is a fixed-size contiguous region of virtual memory backed by 
physical memory from a specific socket node, and each zone can be owned 
by at most one heap instance at a time.
Zones enable: 
(1) locality-aware allocation in big-memory machines with non-uniform 
access latency, 
(2) scalable concurrent allocation of shared memory by multiple processes 
sharing the global memory pool, and 
(3) efficient deallocation of multiple allocated memory blocks at once. 

To allocate memory in a locality-aware manner, a heap finds a zone with 
enough space to satisfy the allocation request that is backed by memory 
from a socket node that matches the location hint.
If the heap cannot find such a zone, the heap tries to acquire a new zone 
backed by memory that matches the location hint, otherwise it tries to 
acquire a zone from a node closest to the location hint. 
To acquire ownership, the allocator locks the zone by atomically writing 
the unique instance number associated with the heap. 

To support variable-size memory allocation, the heap follows a hierarchical 
memory layout that splits zones into variable-size extents, and extents into 
same-size blocks. 
An extent is a contiguous subregion of zone space. 
Its length is a multiple of a configurable page size (default value set to 4KB) 
and can be as small as a page and nearly as large as a zone. 
A extent map at the beginning of the zone is used to track extents by recording
the beginning and length of each extent.  
To speed up locating free extents, each process constructs a private extent-tree 
that tracks per-zone free extents by start address and length.
When the size of the requested chunk is larger than the page size, the 
heap rounds up the allocation request to a page multiple and allocates an extent 
of that size.
If the size of the requested chunk is smaller than the page size, then 
a single-page extent is allocated and formatted as a slab that splits the extent 
into smaller same-size blocks. 
The requested chunk is allocated into the smaller block that it can fit.
A bitmap stored in the slab header is used to mark each allocated block in 
the slab using a bit per block.

The heap supports both freeing an individual memory block and collectively 
freeing all memory blocks allocated through a heap instance.
To free a block, the heap locates the zone containing the block. If the 
zone is owned by the heap instance invoking the deallocation, then the heap 
can free the block. Otherwise, deallocation fails and it is the responsibility 
of the user to direct the free call to the heap and process owning the zone.
To collectively free all memory allocated through a heap instance,
the heap simply zeros all zone metadata of zones that match the 
given heap instance generation number.

The heap supports a limited form of fault tolerance:
the heap guarantees the failure atomicity of individual allocation and free 
operations so that failures during such operations do not corrupt the global 
memory pool, but the heap requires the user to explicitly destroy a heap that 
is no longer needed after a failure so that heap memory can be reclaimed. 
Failure atomicity of individual operations is simplified by the simple 
hierarchical memory layout that allows incrementally turning a zone into 
an extent and an extent into a block using multiple failure-atomic steps. 
Each step is failure atomic as it updates a single memory word, such as 
setting a bit to mark a block as allocated. 

\ignore{
When a process owning a heap fails, The interface also provides support for the fault tolerance necessary in big-data analytic frameworks. When opening a heap, a user can ask for a generation number identifying the current instance of the heap. Blocks associated with a generation number can be released in two ways: (1) explicit call to free the memory block, or (2) a bulk-free call that frees all the blocks associated with a given generation. Deallocating memory through a bulk-free is useful for releasing memory when recovering from a crash.
}

We implement the global memory pool by memory mapping into each user 
process a common shared file that is backed by a memory-resident file 
system such as TMPFS. 
To support locality, zones within the file are preallocated to sockets 
in a round-robin fashion using libnuma.
As each user process may map the shared file at a different location, 
processes cannot use absolute virtual addresses when passing 
references between each other. 
Instead, references use 64-bit offsets relative to the base
of the memory-mapped file. Each process can then create the virtual address
corresponding to the reference by adding the offset to its base 
virtual address. 
To hide this complexity from the programmer, 
we encapsulate the translation process into C++ smart pointers that 
transparently translate the relative offset into a virtual address 
and use the address to refer to the shared memory.

\subsection{Shared-memory Shuffle Engine}
\label{sec:shuffleengine}
Data processing in Spark is divided into stages that compute and communicate data.
Data is ``shuffled'' when re-distributing data from a source ($Map$) stage to a destination ($Reduce$) stage, to support the Spark operators dealing with processing of key/value pairs.
Example operators include $GroupBy$ (to group together values sharing an identical key), $ReduceBy$ (to apply a reduce function on the grouped values sharing an identical key), $PartitionBy$ (to move keys into different partitions), and $SortBy$ (to sort keys with global ordering).

\begin{figure} [h]
  \includegraphics[width=0.5\textwidth]{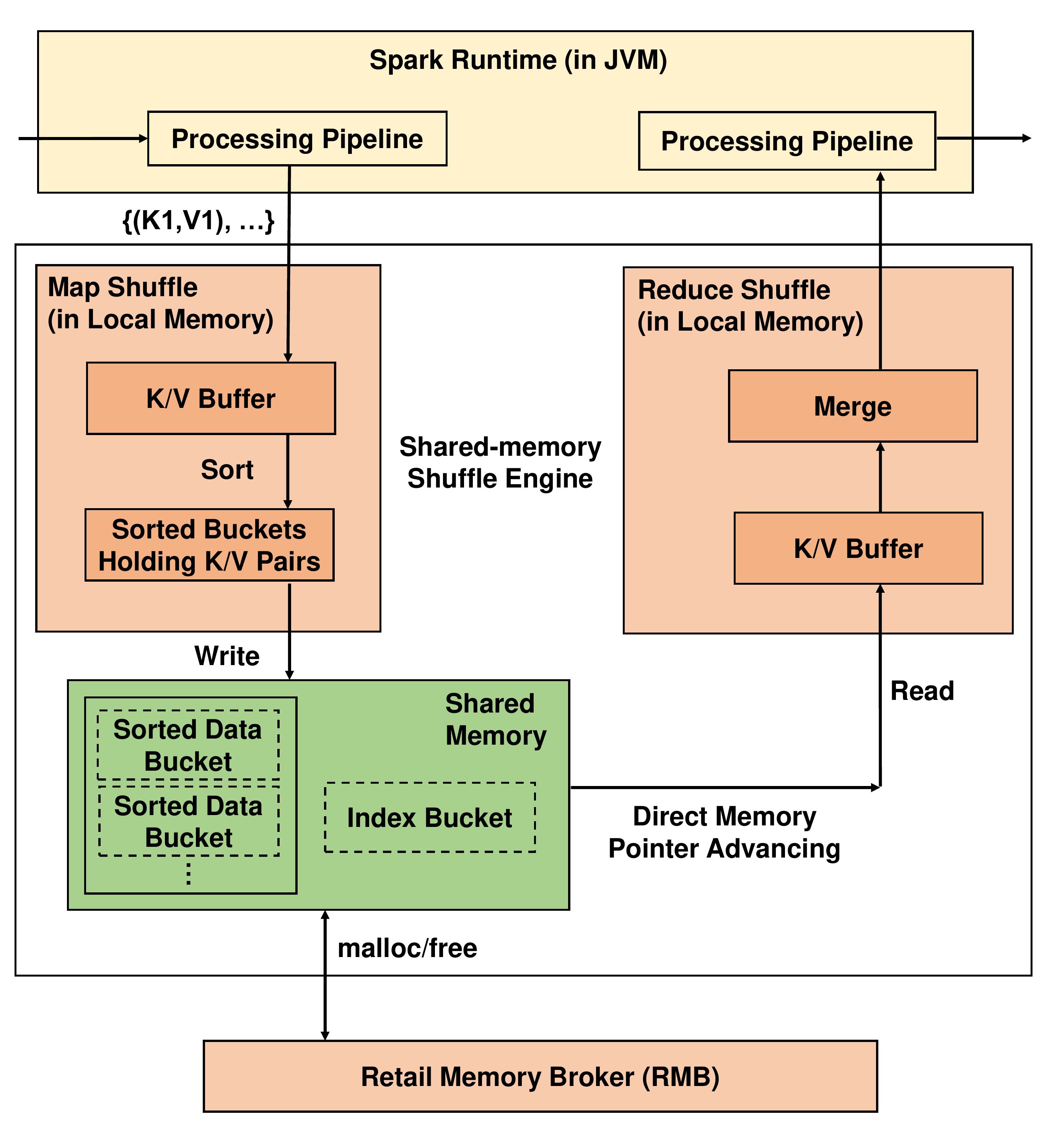}
  \caption{Shared-memory shuffle engine}
  \label{fig:shuffle}
\end{figure}

Figure~\ref{fig:shuffle} shows the architecture of the shared-memory shuffle engine and its integration with the RMB.
The $Map$-side shuffle engine pushes key/value pairs received from the processing pipeline into the internal memory key/value buffer.
The engine then sorts the stored keys and writes the key/value pairs to the shared memory, data bucket by data bucket.
Each data bucket is allocated from the shared-memory region by the RMB.

The $Map$ also writes index information about all of the data buckets into shared memory, as a special index bucket.
A global pointer to the index bucket is sent to the Spark scheduler, which sends it to each destination $Reducer$.
The $Reducer$ uses the global pointer to access the index bucket, from which data buckets can be retrieved by advancing the corresponding global data bucket pointer.
The keys of the data buckets go through the merge engine (e.g., a priority queue), while the corresponding values are stored temporarily in the local memory buffers.
The $Reduce$ processing pipeline stage pulls the keys (the merge engine output) and the merged values (in the local memory buffers) out of the shared-memory shuffle engine.
To improve shuffle performance, we used several techniques.
First, we optimized shuffle for key types compatible to C++ built-in types.
Traditionally in shuffle, both keys and values are stored in the serialized $byte[]$ format, which gets copied and transferred blindly.
Our approach directly writes keys with built-in C++ types (e.g., $int$, $long$, $float$, $double$, $string$ and $byte[]$) to the K/V buffers without serialization.
Keys with arbitrary object types are serialized into $byte[]$.
Second, since each $Map$ or $Reduce$ process may have many concurrent $Map$ or $Reduce$ tasks, we pool and re-use the key/value buffers after the completion of each task.
This technique reduces the total amount of memory required for shuffle and speeds up shuffle dramatically.
Third, we employ different shuffle schemes to separately handle operators that need ordering vs. non-ordered aggregation.
In addition to the sort-based shuffle scheme shown in Figure~\ref{fig:shuffle}, which is used for order-based operators such as $SortBy$, we have developed two additional schemes: hash-map merge and direct pass-through.
The hash-map merge scheme, which performs key/value merging at the Reduce side using hash tables, is used by $GroupBy$ or $ReduceBy$ operators, which need key/value aggregation without ordering.
The direct pass-through scheme, which allows the $Reducer$ to retrieve the $Map$ buckets without ordering or aggregation, is used by $PartitionBy$.

\subsection{Off-heap Memory Store}
\subsubsection{Overview}
\begin{figure} [h]
  \includegraphics[width=0.5\textwidth]{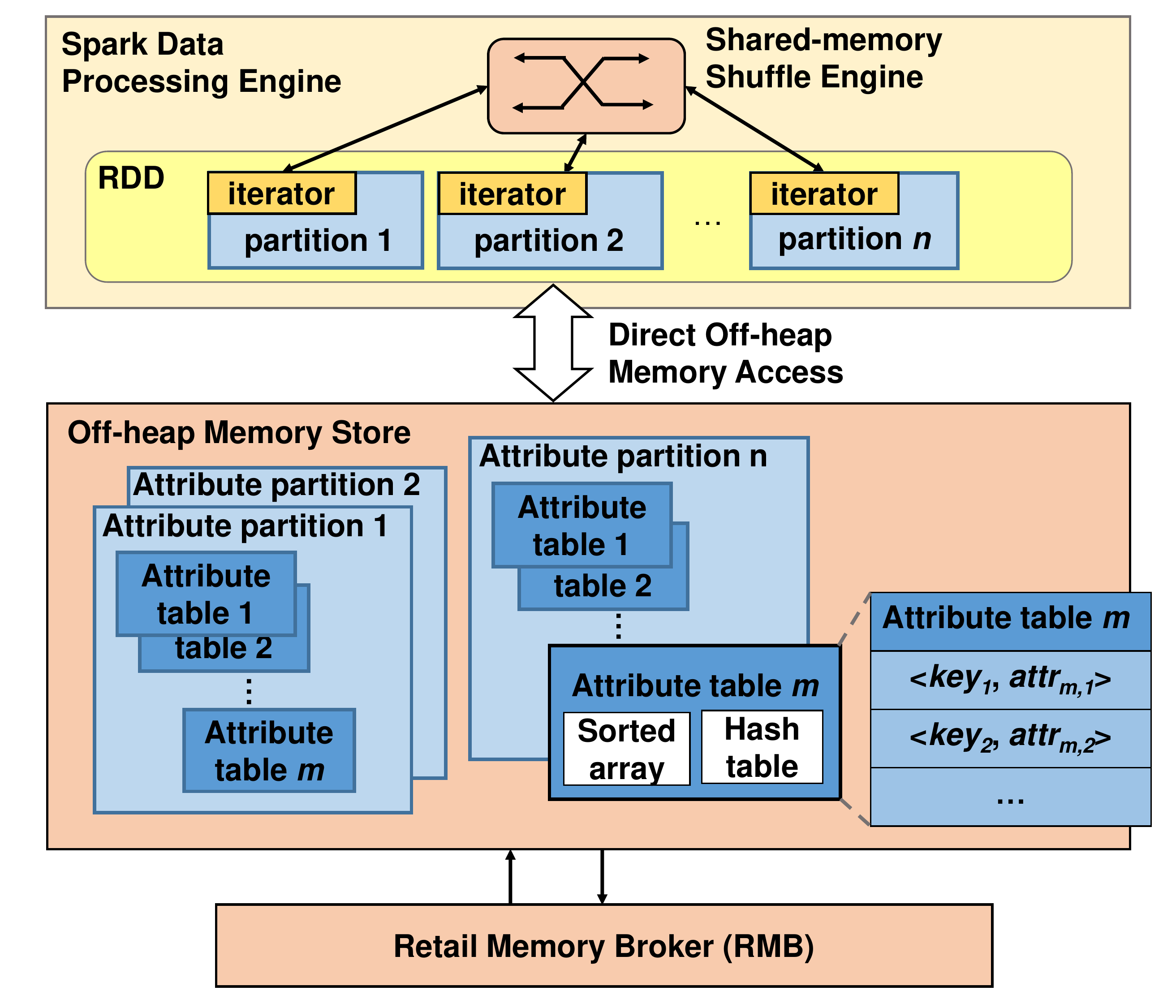}
  \caption{Off-heap memory store}
  \label{fig:offheap}
\end{figure}

Figure~\ref{fig:offheap} shows the overall architecture of the solution based on off-heap memory store that allows us to construct various large-scale data structures in shared-memory regions managed by RMB.

Each RDD consists of partitions and each partition is represented as data structures including sorted array for linear scan and hash table for hash-based search, to store intermediate data processing models, and to allow these models to be updated in place during iteration.

Each RDD is associated with at least one attribute $\{attr_1, attr_2,.., attr_k\}$.
Each RDD partition corresponds to an attribute partition in off-heap memory store. In an attribute partition, each attribute $attr_j$ is allocated an attribute table, which stores as a sorted array and a hash table with pairs of $<key_i, attr_j, i>$.
The off-heap memory store is inherently a distributed partitioned in-memory key/value store that can store and cache attributes for iterative processing.

An RDD partition, considered the owner of the corresponding attribute partition, creates and updates its attributes in each iteration.
Other partitions are allowed read access to attribute partitions they do not own.
In general, such read/write access requires synchronization.
To speed up data processing, partitions often access (i.e., read or write) attributes in bulk.
An RDD partition is constructed and cached at the job initialization phase.
When the job finishes, the executor process terminates along with the cached RDD partition and the associated attribute partition.
Thus, the off-heap memory store for attribute partition is inherently a caching store.

In the shared-memory based architecture supported by the shared memory machine, an RDD partition will have its own attribute partition created locally, i.e., co-located in the same shared-memory region.
An attribute table is constructed via a sorted array data structure, which is laid out contiguously in the process’s address space, in order to support fast bulk access.
For a hash-based partition, direct memory access to a $<key, attr>$ pair is via a pointer to the corresponding shared-memory region.
Subsequent pairs can be accessed by advancing the pointer.

\subsubsection{Globally Accessible Data Structure}
\begin{figure}[h] 
  \includegraphics[width=0.5\textwidth]{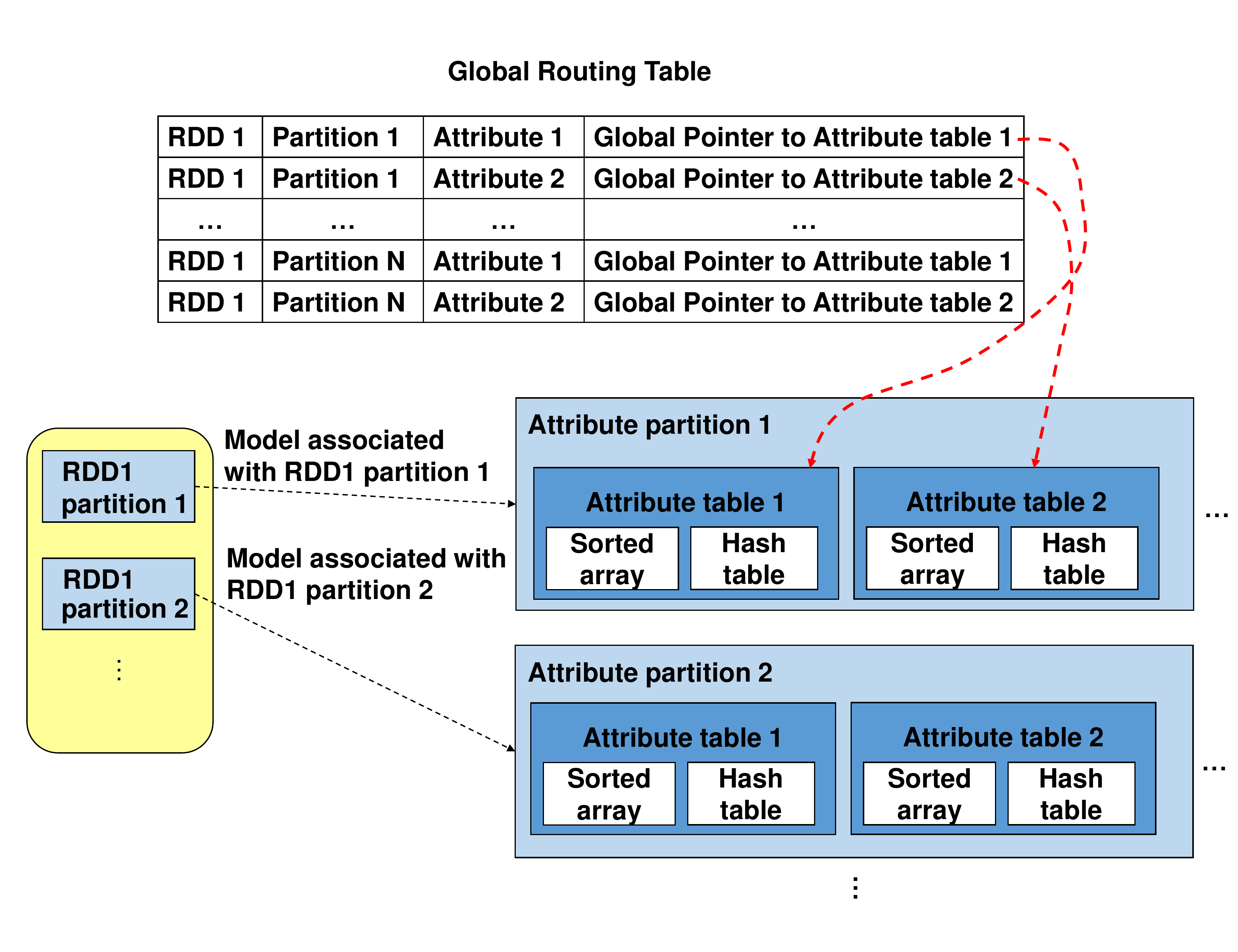}
  \caption{Global routing table for globally accessible attribute partitions}
  \label{fig:routing}
\end{figure}

When each attribute partition associated with an RDD partition is created in the off-heap memory store, its global addresses are collected and stored as a global routing table (see Figure~\ref{fig:routing}).
This global routing table is then broadcast to each Spark executor so that the addresses become globally accessible.

\begin{figure}[h] 
  \includegraphics[width=0.5\textwidth]{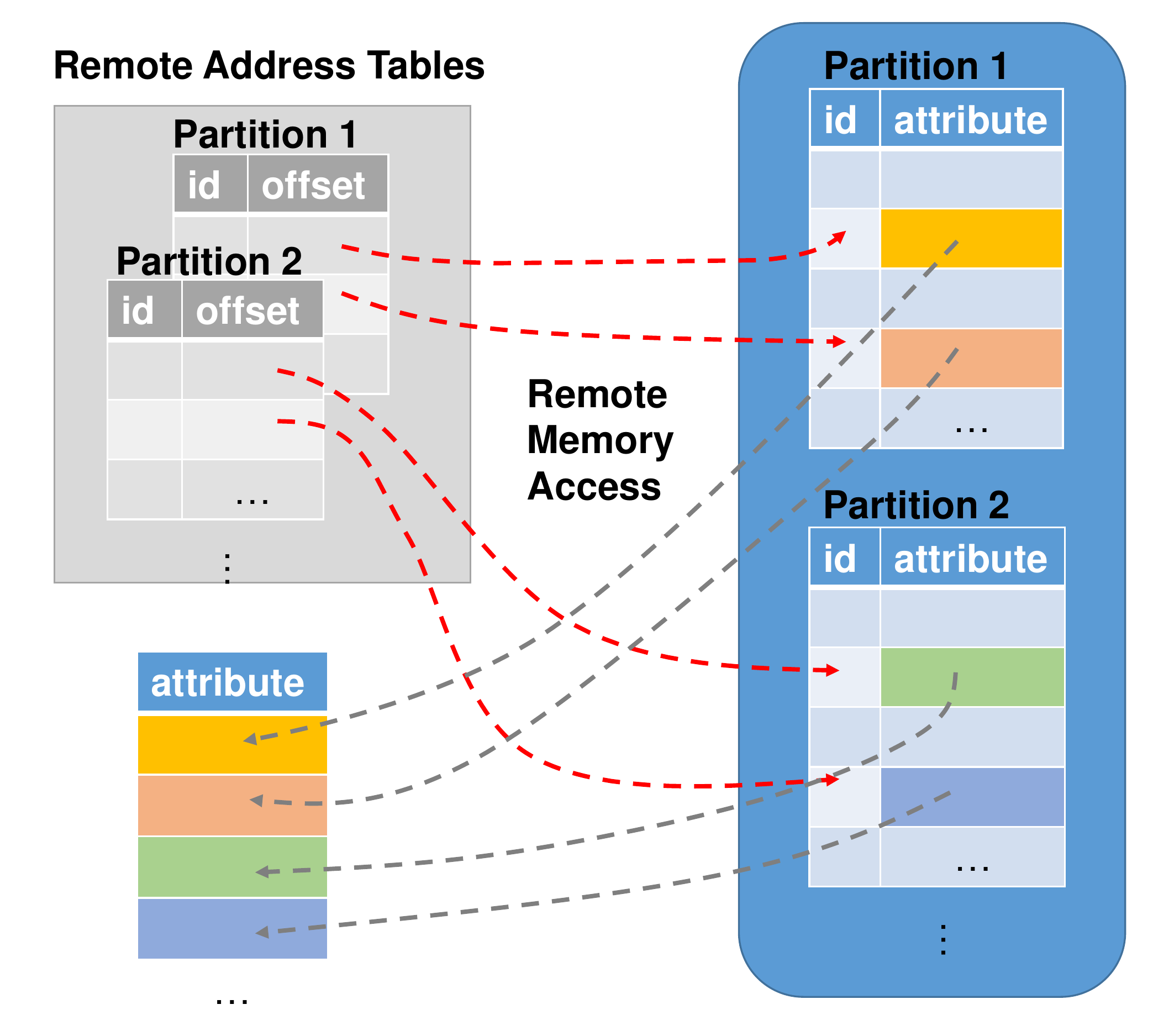}
  \caption{Global address table that contains associated attribute offsets for global memory access}
  \label{fig:address}
\end{figure}

Since these data structures are globally accessible by every Spark executor within the cluster, we can reduce the number of shuffle stages that are required during iterative processing algorithms.
For example, for an edge partition, we create a global address table that contains offsets for the associated vertex attributes.
This global address table is built when the input graph is loaded before the iterative processing.
In each iteration, we retrieve the attribute values for the offsets in the global address table by global memory access as illustrated in Figure~\ref{fig:address}, instead of shipping the vertex attribute values through a shuffle step.

This is achieved by having the data written (or updated) in shared memory in the current processing stage by a single writer which, in turn, is read directly from shared memory by multiple readers in the next processing stage. 
Note that because writing and reading are separated from different Spark stages due to Spark’s bulk-synchronous processing paradigm, there exists no contention from simultaneous writing and reading.

\subsubsection{Fault Tolerance}

Big memory machines are made out of hundreds of independent compute units and other hardware 
components. Thus the overall system mean time between failure (MTBF) is low compared 
to standard MTBF (20 years) of individual components. Therefore applications running long periods
of time (e.g. iterative ML workloads)  on such big memory machines are vulnerable to hardware failures. In addition transient 
errors also contributes to system failures that requires node/ application restarts for recovery.
Worker restarts/ system restarts during such application runs may result in loosing all the compute
progress made till the failure and risks application starting the compute all the way from the 
beginning. Sparkle solves the problem by implementing an efficient checkpoint/ restart
scheme for off-heap store to recover from possible hard/transient failures.

Sparkle provides a 'checkpoint\_partitions()' library call to perform
application level checkpoints. During the library call, each worker dumps the off-heap memory store RDD partition 
states associated with each worker process (partitions it owns) to the stable storage (SSDs) in parallel. 
Parallel applications that confirms  to BSP processing model, often constitute of explicit 
synchronization points that requires all the worker participation (e.g. shuffle step, end of compute stage).
Strategically checkpoint library calls ('checkpoint\_partitions()') are placed before 
such synchronization points to create implicitly synchronized application checkpoints. 
That is during checkpoints each worker, snapshots the RDD partition state residing in the off-heap memory store 
along with a snapshot version number. 

Sparkle implements asynchronous checkpoints. That is during a checkpoint library call, a worker first creates an in-memory
copy of the RDD partitions that it manages  with the steps, (1) allocating new memory buffers that is large enough
to hold partition data (2) acquiring appropriate locks corresponding to the off-heap store partition (3) create a copy of the current
partition state using 'memcpy()' call. After creating a copy of partition data (snapshot), the application is free to 
run ahead, and a background thread copies the created partition snapshot along with a version (we use a monotonically increasing
counter per RDD partition) on to the stable storage (SSDs). By creating an in-memory copy of the partition data we move the 
slow block I/O time during partition snapshots out from the critical path of the application execution. 

During an application restart, the library runtime first scans through partition snapshots stored in stable storage (SSDs) and choose the 
most recent common snapshot version across the all workers for each RDD. Note that the partition snapshots with same version number
across partitions/ workers makes up a consistent state for the corresponding RDD and the all such checkpointed RDDs makes up the 
application consistent state for a application restart. After choosing the set of partitions that makes up the consistent restart state, 
it loads them to each of the workers address spaces and builds up the global routing table before starting the 
computation.

\section{Experiments}
~\label{sec:experiments}
We conducted a series of experiments to estimate the effectiveness of the shared-memory shuffle engine and off-heap memory store.
Our baselines are Vanilla Spark on a scale-out cluster (Vanilla-Scaleout) and Vanilla Spark on a scale-up hardware (Vanilla-Scaleup).

Our experiments include micro-benchmarks ($GroupBy$, $Join$, $PartitionBy$, $ReduceBy$, and $SortBy$ Spark operators) and macro-benchmarks (TeraSort, PageRank and Belief Propagation (BP) applications).
They represent typical Spark benchmarks and workloads.

In Section~\ref{scaleupout}, we show that Sparkle's shared-memory shuffle engine enables micro-benchmark operations, PageRank, and TeraSort benchmark applications to run faster. We compare scale-up and scale-out experiment results for the above applications. Both sets of results show the benefits of running Sparkle on scale-up hardware.

In Section~\ref{dragonhawk}, we compare the performance of Sparkle vs. Vanilla Spark on a large memory machine, using larger data sets.
We also show how the off-heap memory store decreases the memory footprint of the Belief Propagation algorithm on large graphs and enables faster execution.

\subsection{Sparkle vs. Vanilla Spark on Scale-out and Scale-up}
\label{scaleupout}

\subsubsection{Experimental Parameters}
\label{scaleout:setting}


We deploy Sparkle with 8 workers on an HPE Proliant DL580, a scale-up machine with 60 cores and 1.5TB DRAM across four NUMA nodes (sockets). Each worker is assigned 7 cores and 64 GB JVM memory. We used Spark 1.6.1 for the Vanilla Spark configurations in this section. For scale-up experiments, we deploy Vanilla Spark using the same configuration as Sparkle for comparison. For both Sparkle and Vanilla Spark on a scale-up machine, 
we bind each worker process to the CPU and memory of a NUMA node. Vanilla Spark uses TMPFS bound to a NUMA node to store the shuffle data and TCP/IP communication for shuffling, while Sparkle uses the shared-memory shuffle engine.

For the scale-out experiments, we use cluster configurations of 4, 8, 16 and 32 machines. For a fair comparison to scale-up, each Spark worker is set up using 7 cores (same CPU speed) and 64GB JVM memory to match the scale-up machine. 
In Section~\ref{scaleout}, we use a cluster of 4 machines with 56 cores and 512GB JVM memory so that the total resources (in terms of CPU cores and JVM memory size) for the scale-out and scale-up configurations is equivalent. In Section~\ref{scaleout:ru}, we present experimental results running on cluster of 4, 8, 16, 32 machines with the same per-machine settings. We configure our scale-out cluster to use Infiniband~\cite{infiniband2000infiniband} for network communication rather than Ethernet, as this hardware configuration gives us the best scale-out cluster configuration we can achieve in terms of network bandwidth and latency. In Section~\ref{scaleup}, we present experimental results running using the scale-up setting for Sparkle and Vanilla-Spark. 

Our Spark operator micro-benchmarks processed 2 million key-value pairs for $GroupBy$, $Join$, $PartitionBy$ and 4 million key-value pairs for $ReduceBy$ and $SortBy$. 
For the PageRank experiments, we used two different ClueWeb graph data sets~\cite{clueweb}: a 20 GB graph with around 100 million nodes and a 130 GB graph with around 600 million nodes. 
We refer to these data sets as $100M$ and $Full$, respectively. 
For the TeraSort experiments, we used a 256GB input data, with the same amount of intermediate data and sorted output data. 



\subsubsection{Comparing to Scale-out Solution}
\label{scaleout}


In this section, we compare Sparkle performance with Vanilla Spark deployed on scale-out clusters.

\begin{figure} [ht]
	\psfig{figure=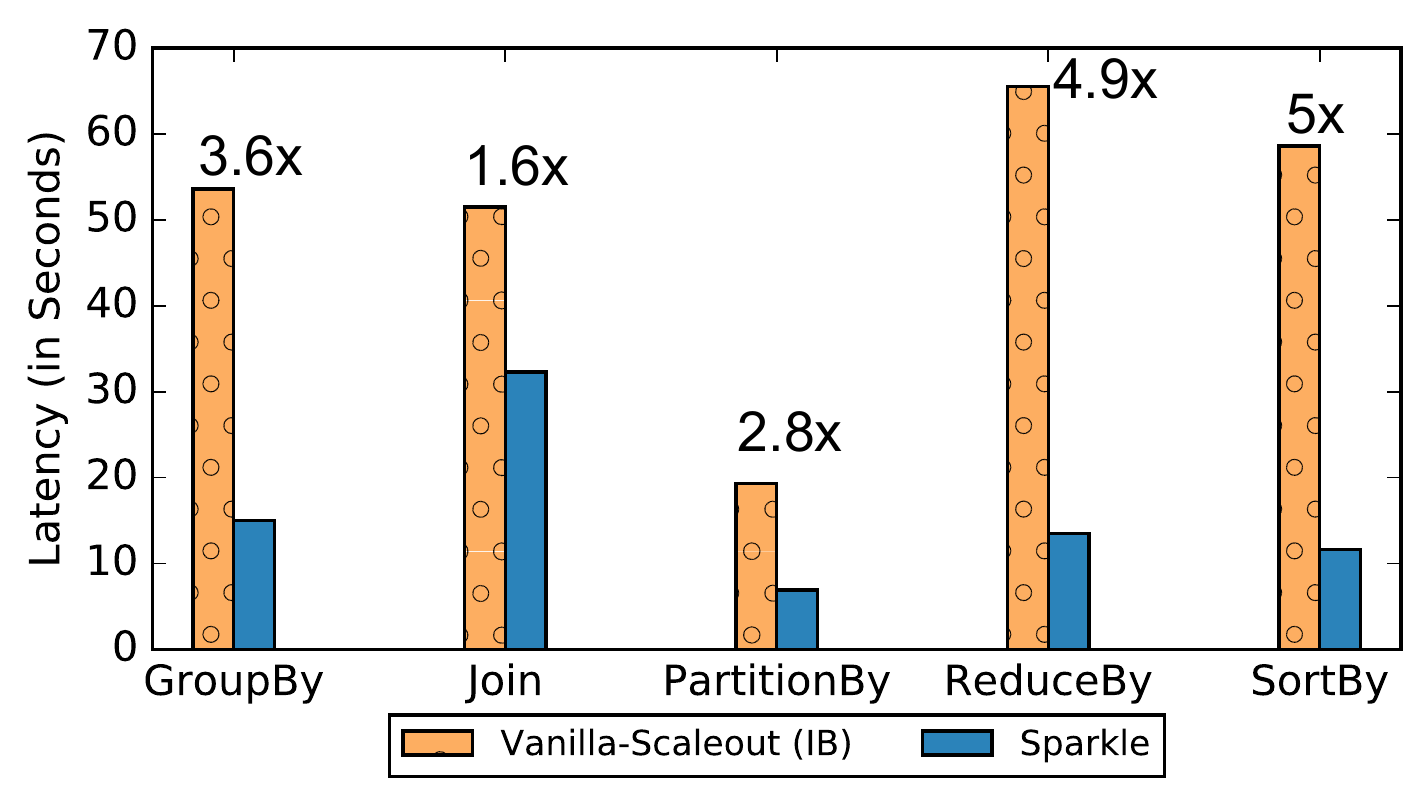, width=3in}
	\caption{Micro-benchmark results with 2M and 4M key-value pairs on Sparkle vs. Vanilla-Scaleout}
	\label{fig:micro_scaleout}
\end{figure}

We ran the micro-benchmarks on Sparkle and Vanilla Spark on scale-out (Vanilla-Scaleout) with TCP/IP over Infiniband (IB). All scale-out experiments are performed using the same number of workers across configurations, where each worker uses the same number of CPU cores and size of JVM memory. The result is presented in Figure~\ref{fig:micro_scaleout}. 

We observe that Sparkle outperforms the scale-out configuration in most cases due to the faster in-memory data access. We also observe slightly less improvement in Sparkle over the equivalent scale-out configuration for $Join$ and $PartitionBy$ operators (1.6x and 2.8x respectively). The computation time is more dominant than the shuffling time in the $Join$ operator and 
Sparkle mainly reduces network and serialization overhead in shuffle. $PartitionBy$ uses the direct pass-through shuffle scheme, which benefits less from the shared-memory shuffle engine than other schemes (see Section~\ref{sec:shuffleengine}). For the rest of the micro-benchmark operators, we observed significant improvements from 3.6x to 5x over the equivalent scale-out setting. 

\begin{figure} [ht]
	\psfig{figure=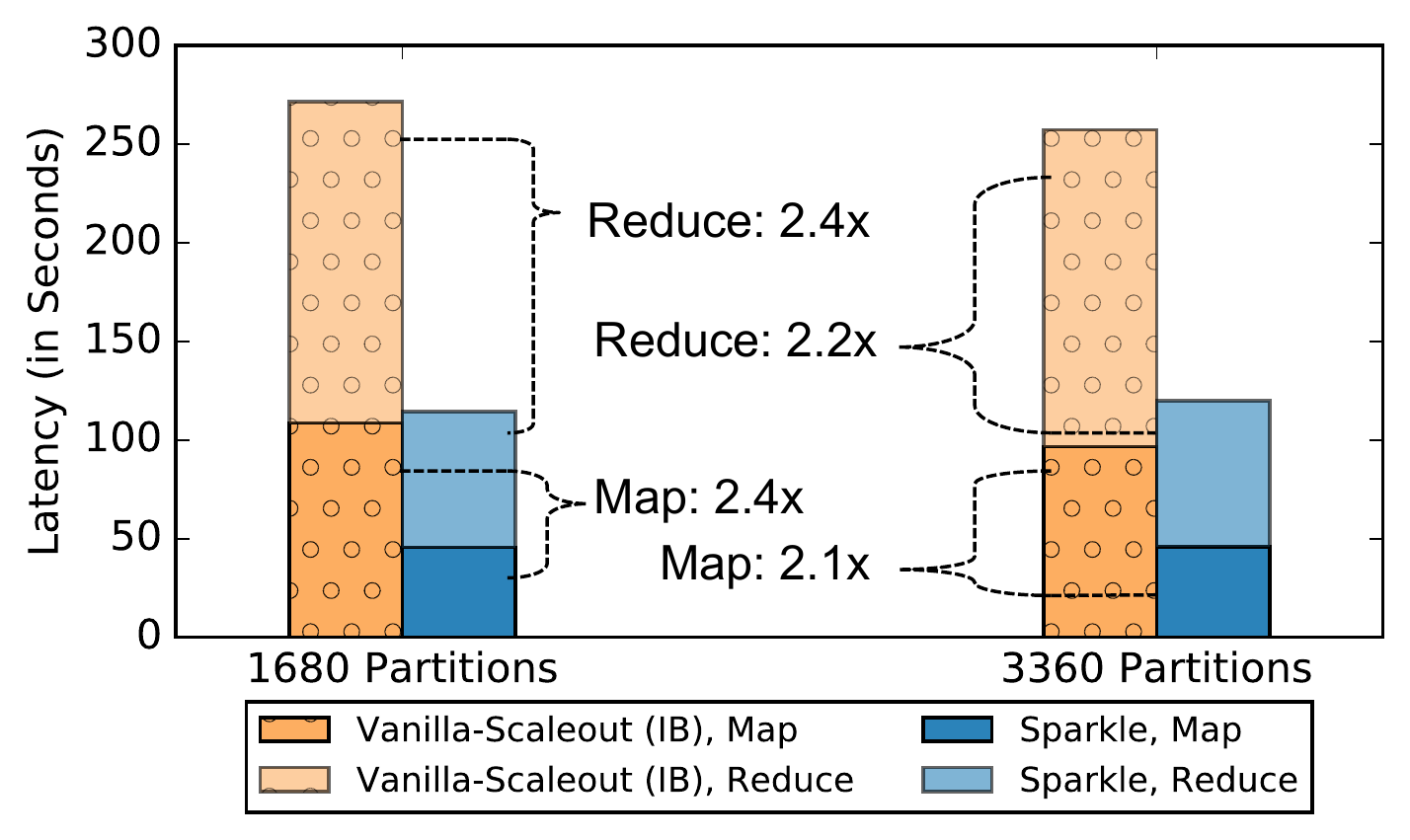, width=3in}
	\caption{TeraSort application on Sparkle vs. Vanilla-Scaleout}
	\label{fig:terasort_scaleout}
\end{figure}
\begin{figure} [ht]
	\psfig{figure=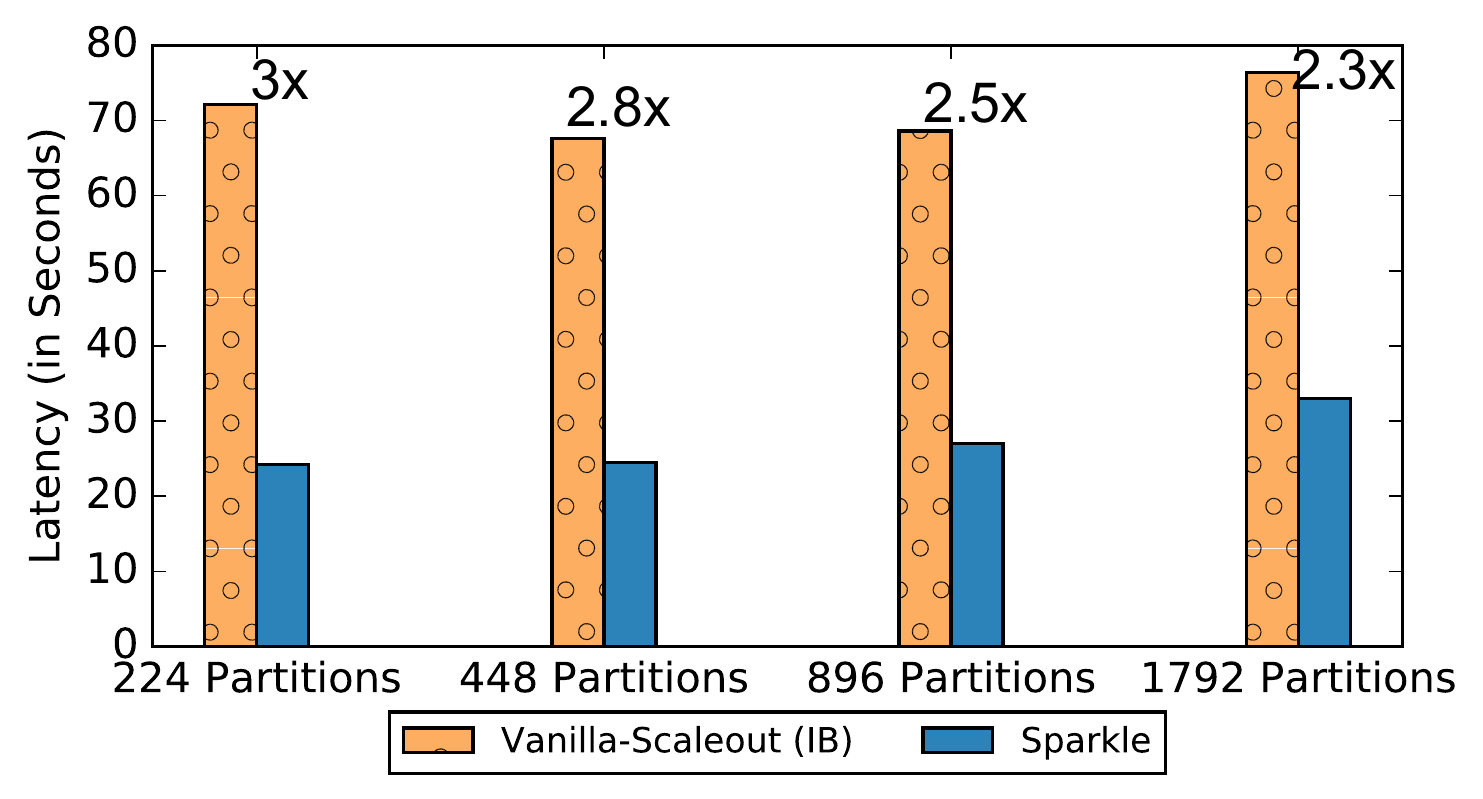, width=3in}
	\caption{PageRank (100M) on Sparkle vs. Vanilla-Scaleout}
	\label{fig:pr_100_scaleout}
\end{figure}

\begin{figure} [ht]
	\psfig{figure=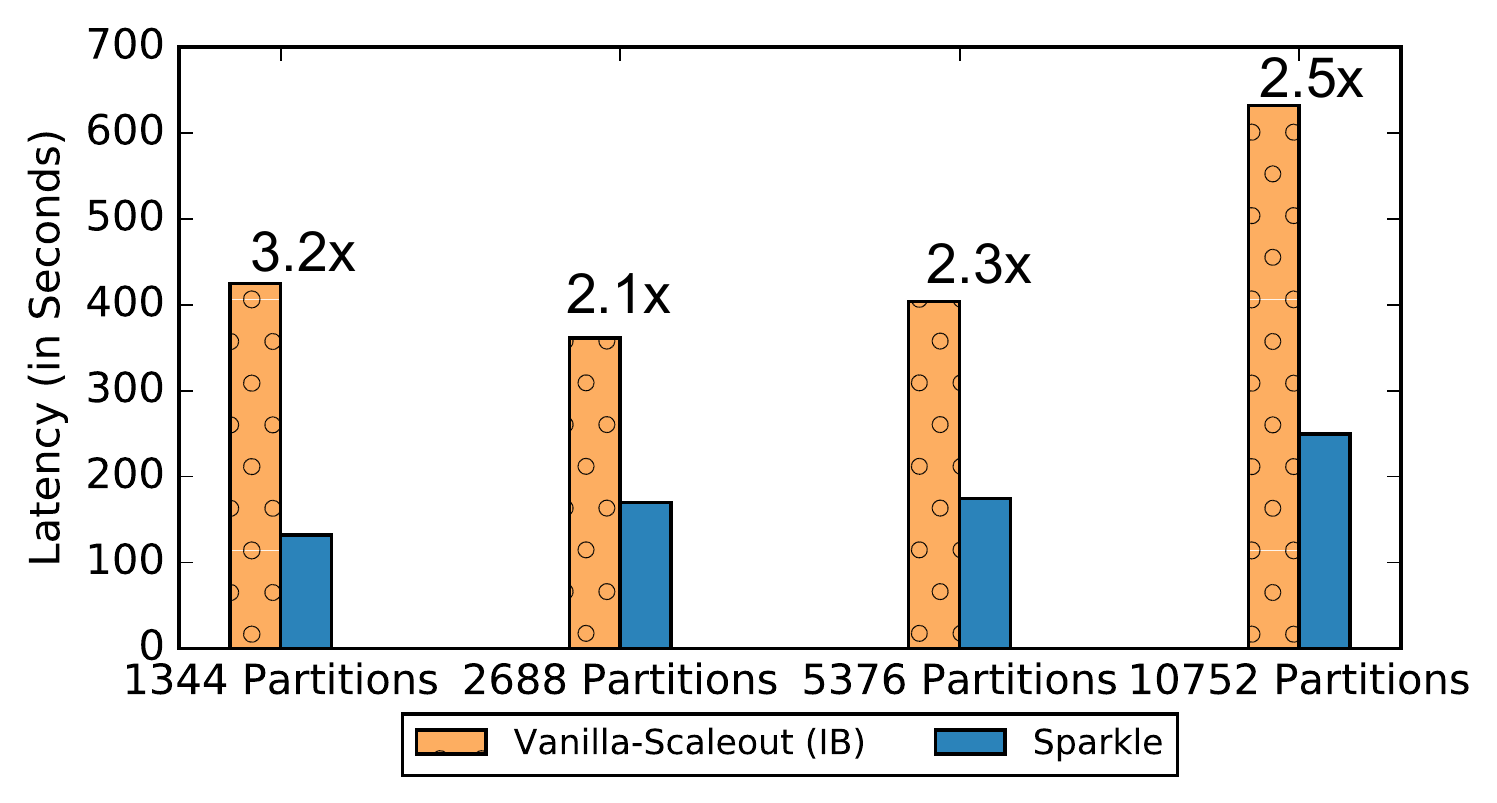, width=3in}
	\caption{PageRank (Full) on Sparkle vs. Vanilla-Scaleout}
	\label{fig:pr_full_scaleout}
\end{figure}

In Figures~\ref{fig:terasort_scaleout},~\ref{fig:pr_100_scaleout} and~\ref{fig:pr_full_scaleout}, we present our Sparkle performance in comparison with Vanilla-Scaleout for TeraSort and PageRank applications.

For TeraSort experiments, from Figure~\ref{fig:terasort_scaleout}, we observe that Sparkle cuts the job completion time in half. The gain is slightly more in fewer partitions. Since TeraSort performs only one shuffle step using $PartitionBy$, less communication bound than computation, Sparkle benefits less from the shared-memory shuffle engine, while the scale-out cluster achieves slightly less overhead in reduce period with larger number of partitions (and hence higher data parallelism). 

For PageRank experiments, we present both 100M and Full data set results. From Figures~\ref{fig:pr_100_scaleout} and~\ref{fig:pr_full_scaleout}, we  observe more than two times performance improvements from Sparkle to Vanilla-Scaleout. Unlike TeraSort results, we observe that the latency slightly increases with the number of partitions for both experiments due to the increasing size of data shuffling by larger number of partitions.  This is because PageRank involves several shuffle steps with $ReduceBy$, which is more communication bound. Moreover $ReduceBy$ has more gain than $PartitionBy$ as shown in Figure~\ref{fig:micro_scaleout}, which makes PageRank using $ReduceBy$ has better gain for Sparkle than TeraSort using $PartitionBy$. 

\subsubsection{Resource Usage Comparison for Sparkle and Vanilla-Scaleout}
~\label{scaleout:ru}
\begin{figure} [ht]
	\psfig{figure=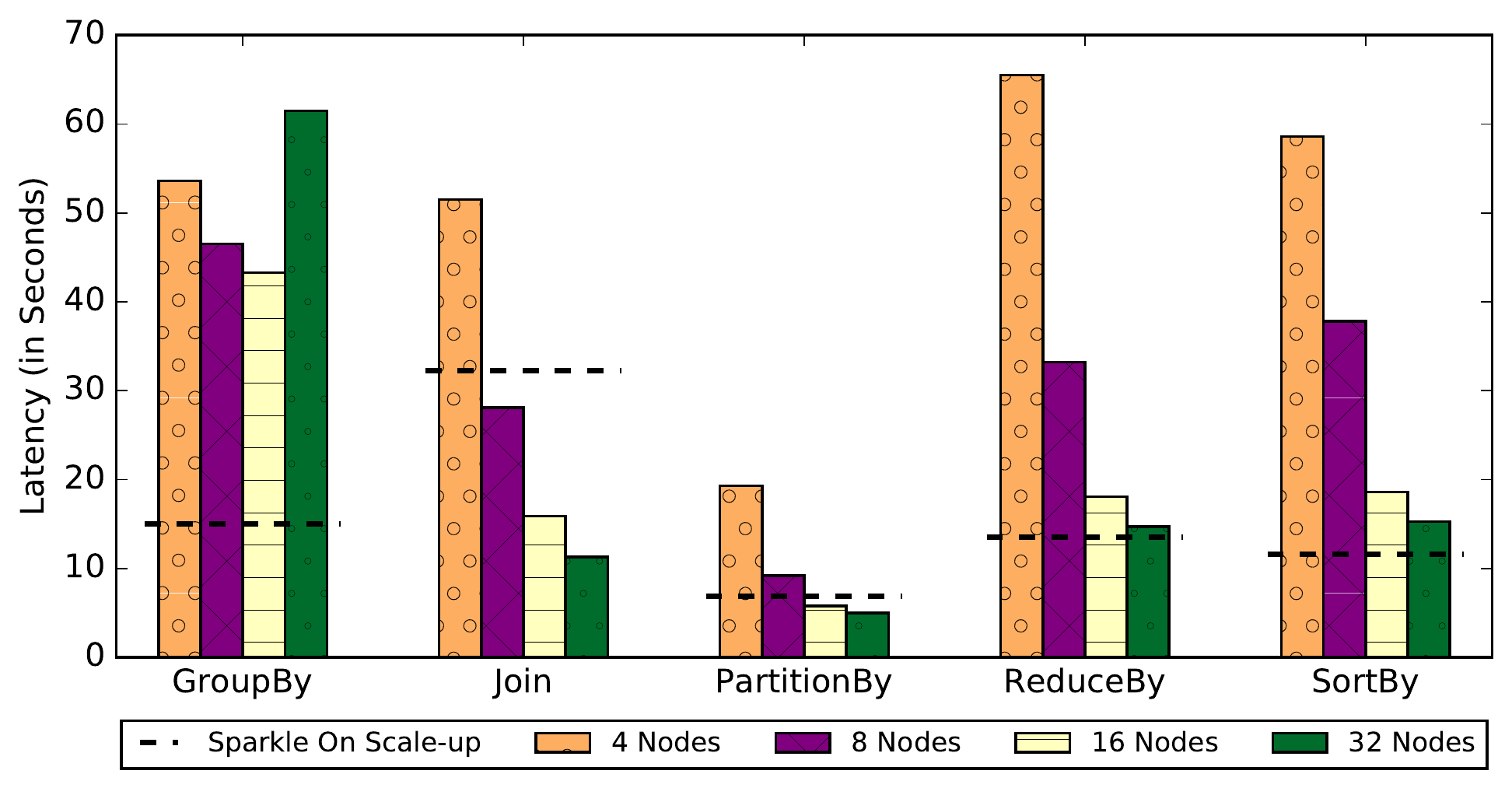, width=3in}
	\caption{Resource usage comparison for micro-benchmarks}
	\label{fig:resource_compare}
\end{figure}

We increase the size of cluster to match the Sparkle performance. We present the results of micro-benchmarks deployed on 4-, 8-, 16-, and 32-node clusters in Figure~\ref{fig:resource_compare}. Among these clusters the 4-node cluster has the same total hardware usage as Sparkle on a scale-up machine (as discussed in Section~\ref{scaleout:setting}). In this figure, the dotted line denotes the job completion time achived by Sparkle and the bars denote the job completion times of Vanilla-Scaleout.
 
From Figure~\ref{fig:resource_compare}, we observe that starting from the equivalent hardware resource (4 nodes), scale-out cluster needs to increase its resource usage by 2 times ($Join$), 4 times ($PartitionBy$) and 8 times ($ReduceBy$) respectively, to achieve shorter job latencies than Sparkle on the scale-up. For $GroupBy$ and $SortBy$ experiments, Sparkle achieves a better job completion time than Vanilla-Scaleout on 32-node cluster (8 times resource usage comparing to Sparkle). Interestingly, we observe for $GroupBy$ operator, a larger cluster does not lead to a better performance. The 32-node cluster results in worse performance than the 4-node cluster. As discussed in Section~\ref{scaleout} , the larger number of partitions introduces more shuffling data, especially for communication bound workloads, which starts becomes a major component in increasing the latency, and thus leads to a longer finishing time. Meanwhile Sparkle achieves a better performance by saving the network/serialization overhead without generating more partitions. This shows Sparkle's advantage towards operators with heavy partition overhead, which cannot be benefited by higher parallelism level.

\subsubsection{Comparing to Scale-up Solution}
\label{scaleup}

\begin{figure} [t]
	\psfig{figure=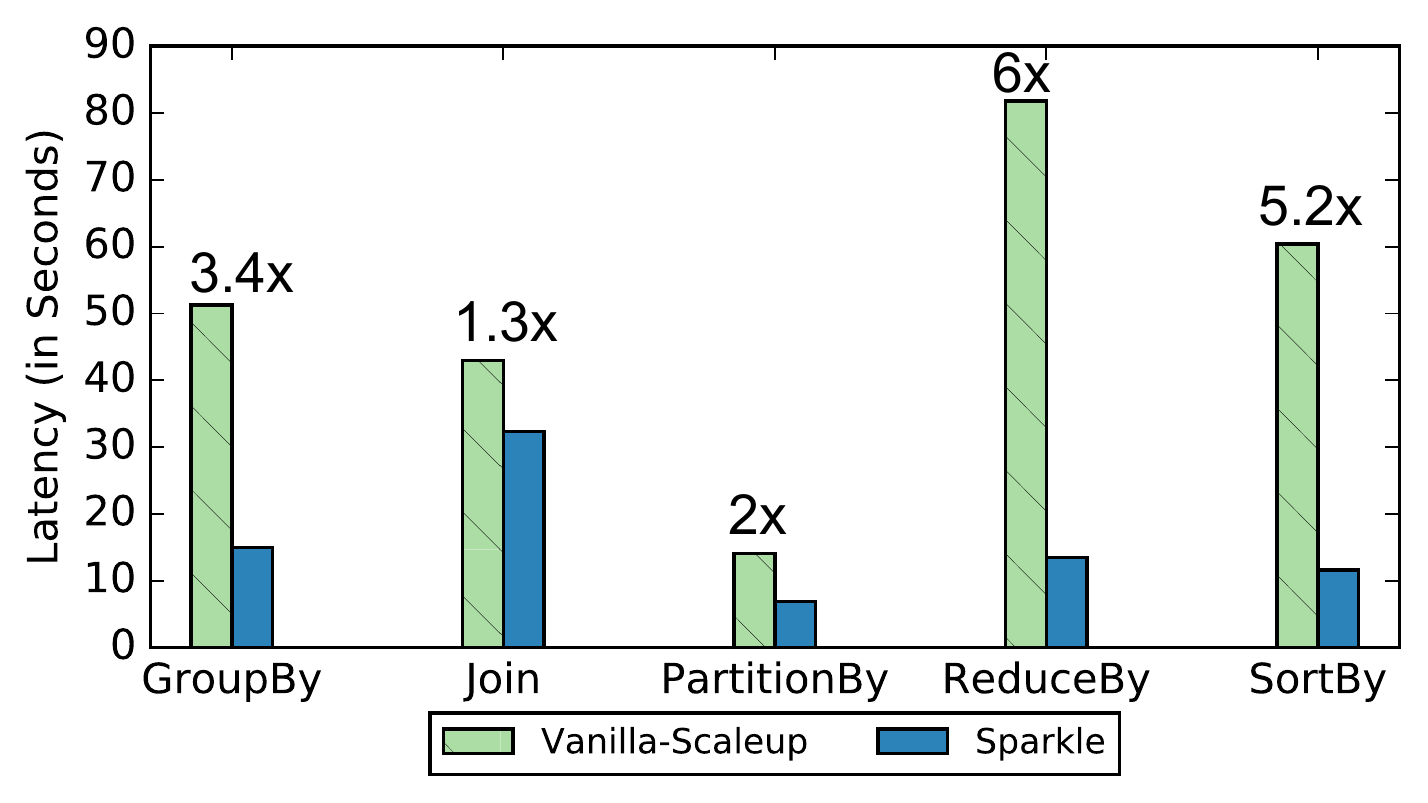, width=3in}
	\caption{Micro-benchmark results with 2M and 4M key-value pairs on Sparkle vs. Vanilla-Scaleup}
	\label{fig:micro_scaleup}
\end{figure}
For Sparkle vs. Vanilla-Scaleup, we observed overall the similar results of the scale-out experiments shown eariler.
As presented in Figure~\ref{fig:micro_scaleup}, Sparkle achieves significantly higher improvement on Vanilla Spark for network-bound operators (e.g. $ReduceBy$ with 6 times improvement) while for compute-bound operators such as $Join$, Sparkle has less gain however it can still manage to achive 33\% faster job running time than Vanilla Spark.

\begin{figure} [ht]
	\psfig{figure=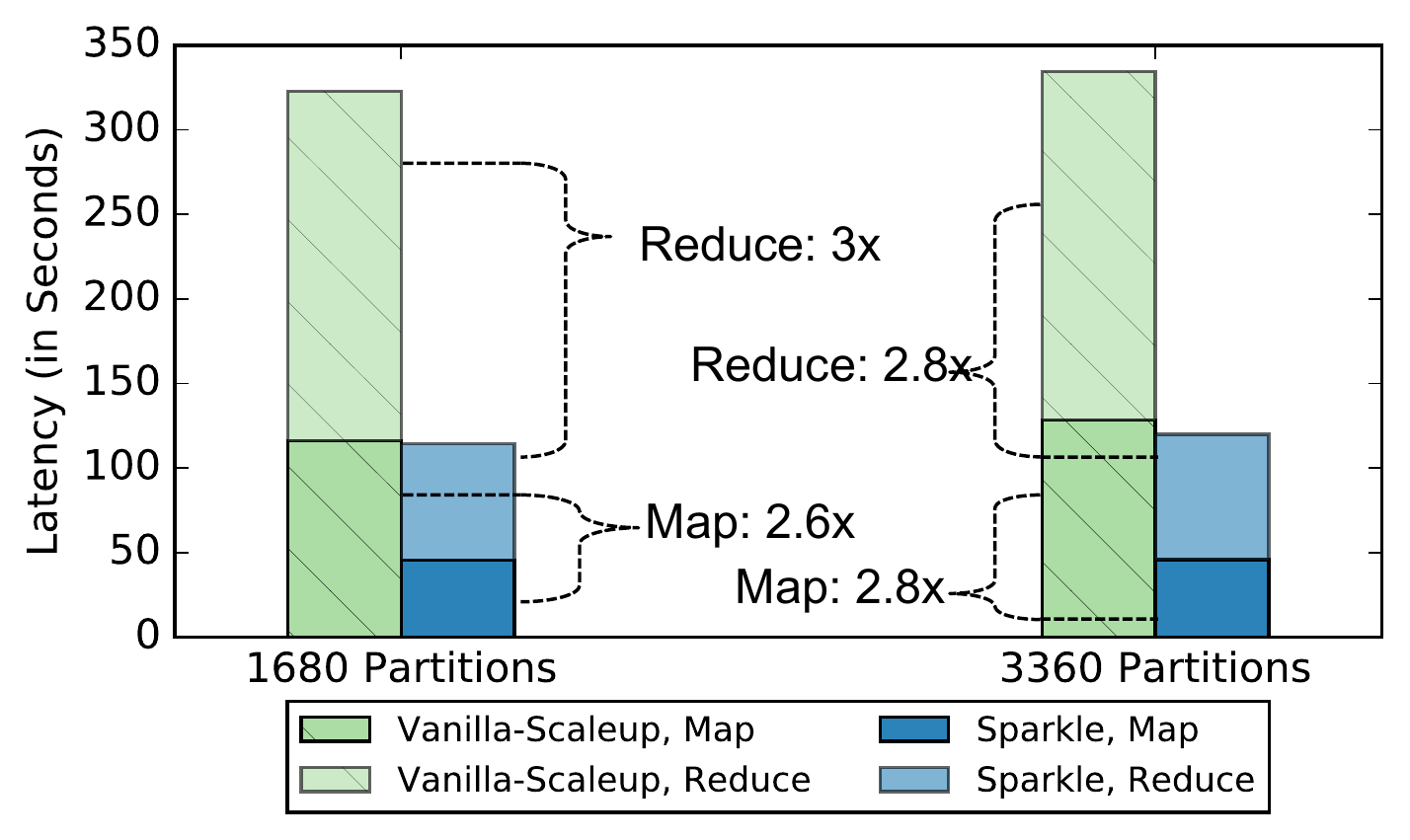, width=3in}
	\caption{TeraSort on Sparkle vs. Vanilla-Scaleup}
	\label{fig:terasort_scaleup}
\end{figure}
For TeraSort experiment in Figure~\ref{fig:terasort_scaleup}, we observe Sparkle improves both map stage and reduce stage significantly, with 2.6x improvement for map and 2.8x improvement for reduce while using 1680 partitions, and 3x improvement for map and 2.8x improvement for reduce. 
\begin{figure} [ht]
	\psfig{figure=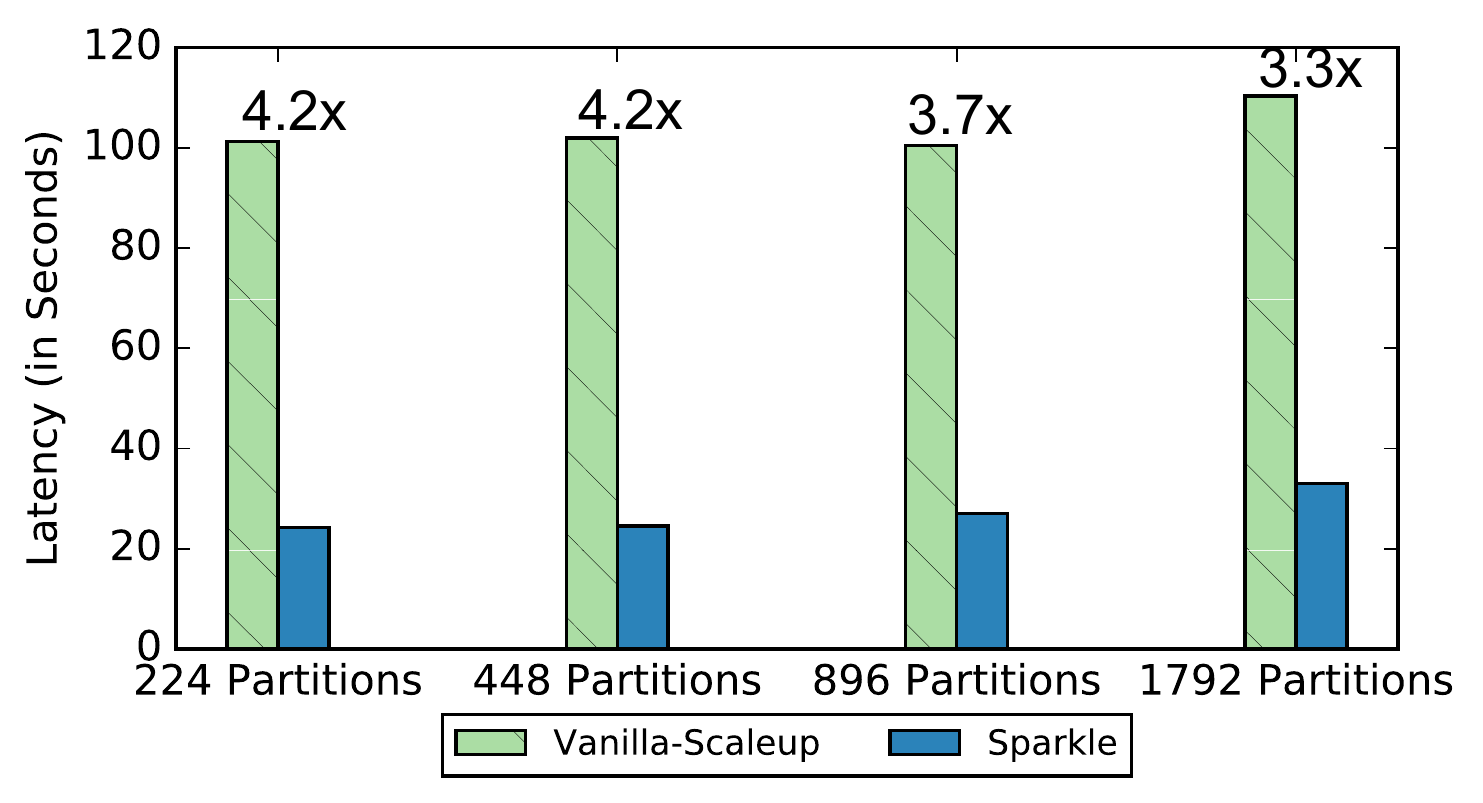, width=3in}
	\caption{PageRank (100M) on Sparkle vs. Vanilla-Scaleup}
	\label{fig:scaleup_100M}
\end{figure}

\begin{figure} [ht]
	\psfig{figure=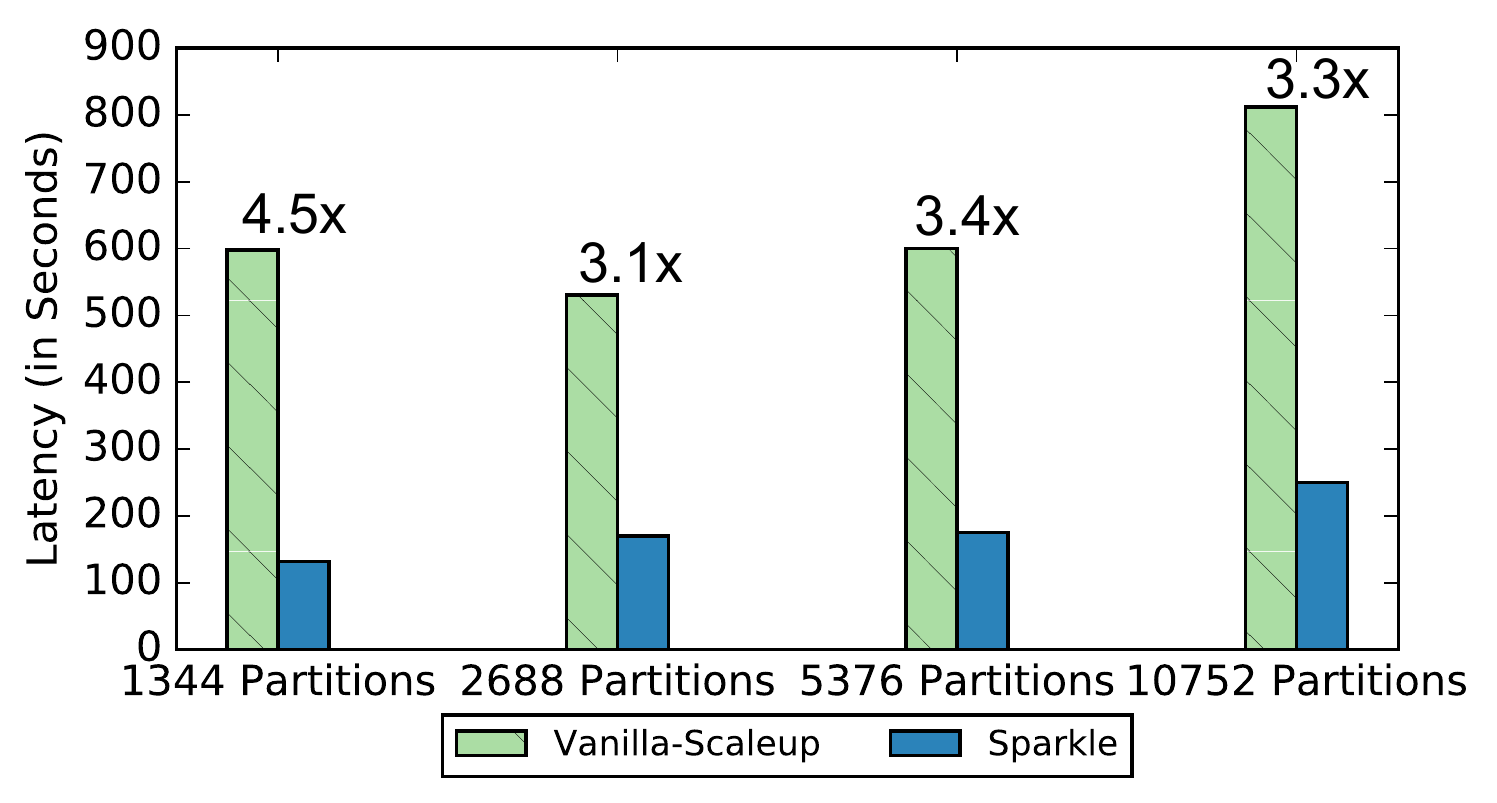, width=3in}
	\caption{PageRank (Full) on Sparkle vs. Vanilla-Scaleup}
	\label{fig:scaleup_full}
\end{figure}

For PageRank experiment, as shown in Figures~\ref{fig:scaleup_100M} and~\ref{fig:scaleup_full}. Sparkle generally achieves 3x to 4.5x better performance than Vanilla-Scaleup. Similarly to the scale-out results in Section~\ref{scaleout}, we found the latency for Vanilla-Scaleup increases with the number of partitions: The partition overhead starts to become significant when we use 10752 partitions. 

\subsection{Sparkle vs. Vanilla Spark on Large Memory Machine}
\label{dragonhawk}

We further performed our experiments with larger data sets to evaluate Sparkle vs. Vanilla Spark on a large memory machine.
Our large memory platform is the Superdome X with 240 cores and 12 TB DRAM across 16 NUMA nodes (sockets).
Each worker process is bound to the CPU and memory of a NUMA node.
For Vanilla Spark, we used TMPFS bound to a NUMA node to store the shuffle data and TCP/IP communication for shuffling. Sparkle uses the shared-memory shuffle engine.
We used Spark 1.2.0 with 45 executors each configured with 96 GB of JVM memory.

We tested micro-benchmarks processing 5.4B key/value pairs in both Sparkle and Vanilla Spark. Keys and values are C++ built-in data types.

\begin{figure} [t]
	\psfig{figure=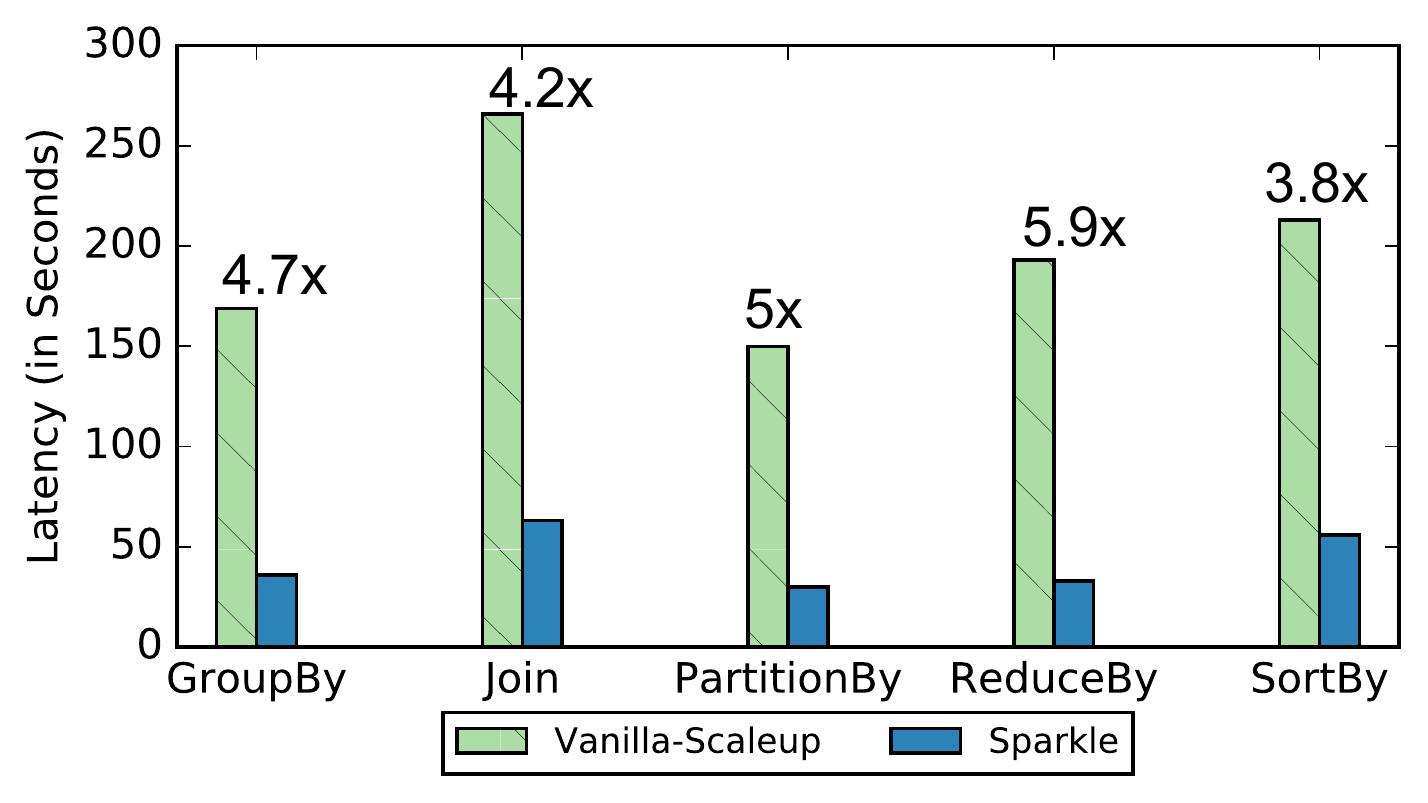, width=3in}
	\caption{Micro-benchmark results with 5.4B key/value pairs on Superdome X}
	\label{fig:dh}
\end{figure}

Figure~\ref{fig:dh} shows the execution times of Sparkle and Vanilla Spark on 5.4B key/value pairs. The gains are from 3.8x to 5.9x, which are overall similar to those with smaller data sets shown in Section~\ref{scaleupout}. 




\begin{figure} [t]
\psfig{figure=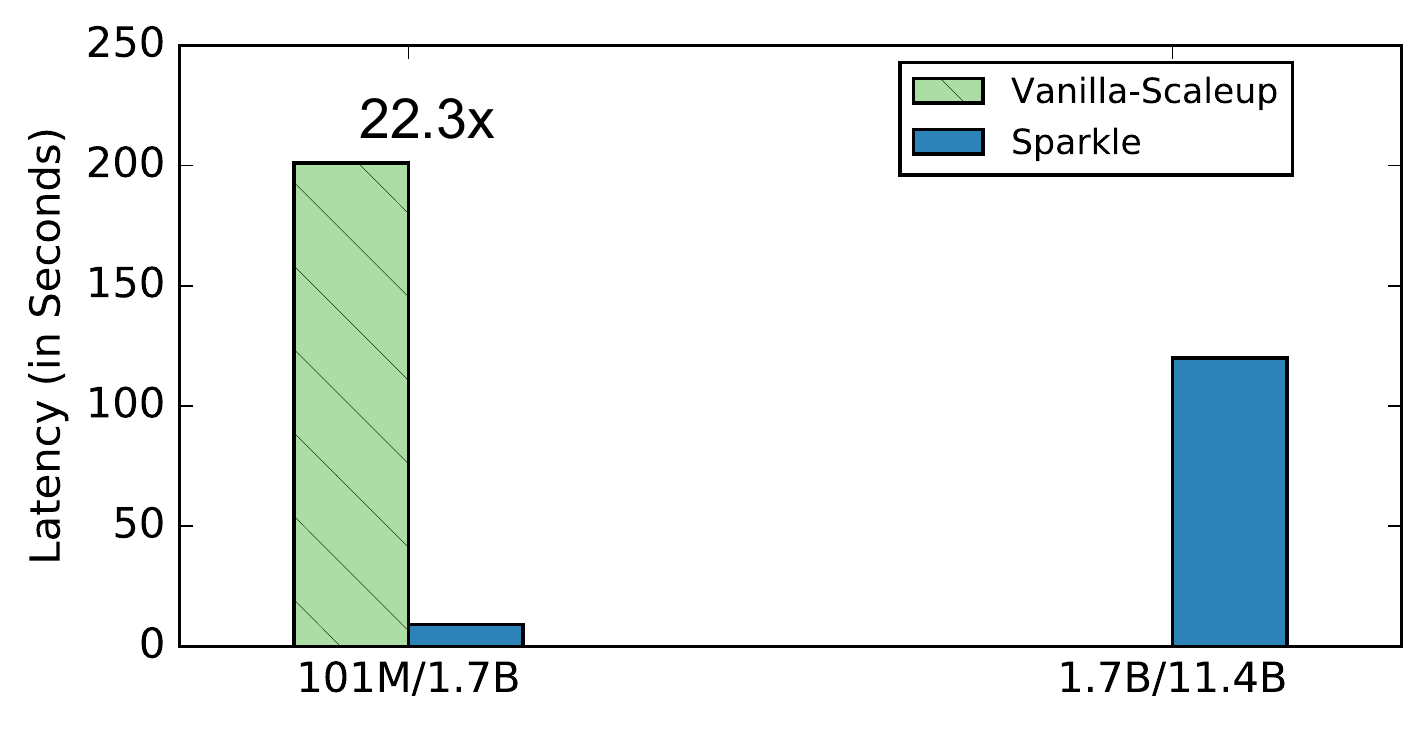, width=3in}
\caption{BP result on Sparkle vs. Vanilla Spark}
  \label{fig:bp}
\end{figure}

\begin{table}[t]
  \centering
  \caption{Memory used for BP}
  \label{tab:memoryuse}
  \begin{tabular}{ | c | c | c | }
    \hline
    Graph size & \multirow{2}{*}{Sparkle} & \multirow{2}{*}{Vanilla Spark}  \\ 
   (nodes/edges) &  & \\ \hline \hline
    101M/1.7B & $\sim$144GB & $>$700GB  \\ \hline
    1.7B/11.4B & $\sim$1.1TB & Fails to run \\ \hline
  \end{tabular}
\end{table}

Finally we demonstrate graph processing for Sparkle with the off-heap memory store vs. Vanilla Spark. We used two large-scale graph workloads: 1) a web
graph with 101 M nodes /1.7 B edges, and 2) a synthetic
graph based on DNS data with 1.7 B nodes / 11.4 B
edges. We can only run these workloads on Superdome X since these workloads consume lots of memory (see Table~\ref{tab:memoryuse}). 

We chose belief propagation (BP) as a representative graph processing workload.
BP is an iterative, message passing algorithm to perform inference on probabilistic graphical models.
Each vertex is associated with a belief vector, and edges are associated with messages.
In each iteration, a vertex aggregates all incoming messages, computes beliefs and computes new outgoing messages.
New messages are computed using the current belief and the messages from the previous step.
This increases the space complexity of the algorithm comparing with PageRank, which does not require storing messages.
BP terminates when all messages have converged.
We also implemented another BP algorithm leveraging Pregel API which supports basic graph processing (send, aggregate edge messages, and modify vertex attributes) in Spark GraphX. We used this Pregel-based BP as our baseline for Vanilla Spark.

In Sparkle BP,
we store graph attributes in the off-heap memory store.
During each iteration, attributes are retrieved from the store, computed and written back to the store through direct off-heap memory access.
The attributes are aggregated and the aggregated results are redistributed via the shared-memory shuffle engine.
The data structures developed, including sorted array and hash table, store intermediate data processing models thus allowing for updates-in-place during iteration.

To further
speed up graph processing in Sparkle BP, we developed a static
graph partitioning mechanism to reduce vertex duplication
across partitions, and offloaded the most computationally
heavy routines (based on profiling) from Scala to
C++.

As seen in Figure~\ref{fig:bp}, for the first workload, Sparkle BP achieves $>$20x speed-up compared to BP with Vanilla Spark.
The off-heap memory store contributes 4x, while the remaining speed-up mostly results from run-time support from the shared-memory shuffle engine and RMB.
The average iteration time is about 13 sec., and it converges in about 10 iterations.
The memory requirement is also reduced compared to the baseline: 144 GB off-heap without on-heap cache for Sparkle vs. more than 700 GB for Vanilla Spark (see Table~\ref{tab:memoryuse}).

We further optimized the Sparkle BP by taking advantage of the globally visible data structures in the off-heap memory store, by consolidating two shuffle stages (out of the total three stages in one iteration) into one single stage.
These three shuffle stages are 1) the edge messages are aggregated, 2) the updated beliefs are reduced to find the maximum difference from the previous beliefs for convergence test, and 3) the updated beliefs are redistributed to the edge partitions.
Here, the 3) shuffle stage is done through global memory access.
Such optimization allows us to further reduce the per-iteration time of 13 sec. to 9 sec.

To further test scalability, we also ran the second workload (11.4 B edges), which is ~2 min per iteration.
The Vanilla Spark BP fails to run at this scale (see Figure~\ref{fig:bp}).

\begin{figure} [t]
	\psfig{figure=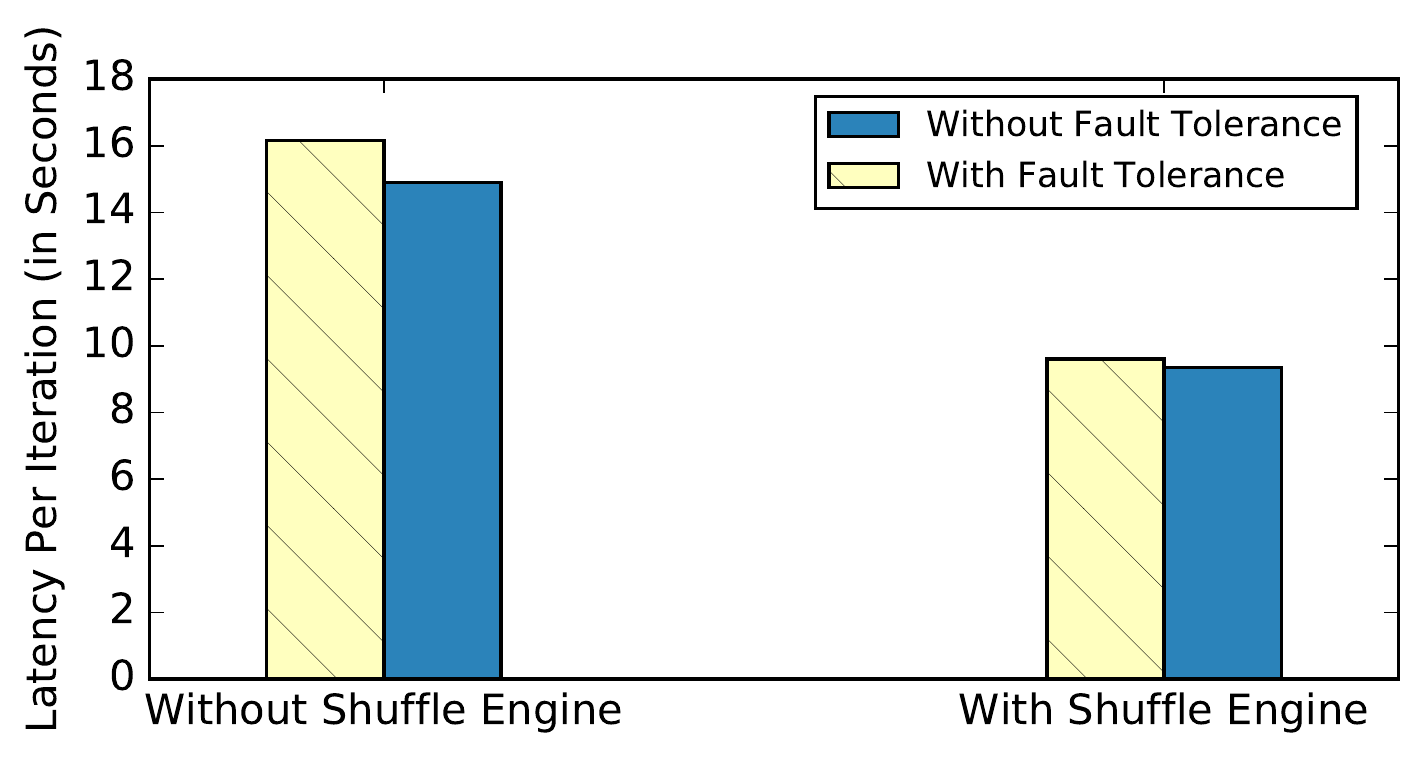, width=3in}
	\caption{Fault tolerance comparison for BP}
	\label{fig:ft}
\end{figure}

We finally quantify the fault tolerance performance for the off-heap memory store by running checkpoint enabled version of the BP application on the first workload (101 M nodes /1.7 B edges). We aggressively insert checkpoint calls after each shuffle step of the BP application to simulate the
worst case checkpoint overhead associated with the BP algorithm. Figure~\ref{fig:ft} shows that there is no significant difference between the checkpoint enabled BP for fault tolerance and the BP without fault tolerance in terms of the iteration time in both cases with and without shared-memory shuffle engine. When using the shared-memory shuffle engine, the gap gets even less.
$\sim$8\% execution time overhead compared its no checkpoint counterpart, in the worst case -- thus confirming the 
efficiency of Sparkle fault tolerance implementation.

In this paper, our experiments have been done in Spark 1.6.1 and 1.2.0. We also experimented with a later Spark version (e.g., 2.0) for some of our workloads and obtained more or less similar gain for Sparkle. We expect that our Sparkle performance is not affected much for a different Spark version.

\section{Conclusion}
\label{sec:conclusions}

We have presented Sparkle, an enhancement of Spark that optimizes 
its performance on large memory machines for memory and communication 
intensive and iterative workloads. Comparisons to a scale-out cluster 
configuration show that a scale-up approach can achieve better 
performance for the same amount of memory and number of cores
due to faster communication between compute nodes.

We have released Sparkle, our shuffle engine and off-heap memory store code to the public under Apache 2.0 License~\cite{sparkle}. We have also released the generalized version of belief propagation algorithm~\cite{sandpiper}.

\section*{Acknowledgments}
We thank Carlos Zubieta for performing preprocessing of the graph used for belief propagation experiments. We also thank Tere Gonzales and Janneth Rivera for working on Spark Demo UI for HP Discover in December 2015.

\bibliographystyle{abbrv}
\bibliography{template}

\end{document}